\documentclass[a4paper,fleqn]{cas-sc}

\definecolor{cite_color}{rgb}{0.0, 0.58, 0.71}
\usepackage{natbib}
\setcitestyle{authoryear,open={},close={}} 
\usepackage{hyperref}
\hypersetup{colorlinks=true,citecolor=cite_color,linkcolor=cite_color}
\usepackage{amsthm}
\usepackage{subcaption}
\usepackage{lineno}
\usepackage{xcolor}
\definecolor{db}{rgb}{0.0, 0.2, 0.7}

\renewcommand{\figurename}{Fig.}

\makeatletter
\renewcommand*{\fnum@figure}[1]{\figurename~\thefigure.}
\makeatother
\def\tsc#1{\csdef{#1}{\textsc{\lowercase{#1}}\xspace}}
\tsc{WGM}
\tsc{QE}
\tsc{EP}
\tsc{PMS}
\tsc{BEC}
\tsc{DE}

\begin{document}
\let\WriteBookmarks\relax
\def\floatpagepagefraction{1}
\let\printorcid\relax 

\def\textpagefraction{.001}
\shorttitle{}
\shortauthors{Mohammad Anis et~al.}

\title [mode = title]{ Pedestrian crash causation analysis near bus stops: Insights from random parameters Negative Binomial-Lindley model}

\author[1]{Mohammad Anis}[orcid=0000-0002-7878-6352]
\cormark[1]
\ead{mohammad.anis@tamu.edu}
\credit{Conceptualization, Methodology, Writing – original draft,  Software, Writing – review \& editing}

\address[1]{Zachry Department of Civil $\&$ Environmental Engineering, Texas A$\&$M University, College Station, TX 77843, USA}

\author[2]{\textcolor{black}{Srinivas R. Geedipally}}[]
\credit{Conceptualization, Methodology, Writing – review \& editing, Supervision}

\address[2]{Center for Transportation Safety, Texas A$\&$M  Transportation Institute, 111 RELLIS Parkway
Bryan, TX 77807, USA
}

\author%
[1]
{\textcolor{black}{Dominique Lord}}
\credit{Conceptualization, Methodology, Writing – review \& editing, Supervision}

\cortext[cor1]{Corresponding author}

\begin{abstract}
Pedestrian safety remains a pressing concern near bus stops along urban transit, where frequent pedestrian-vehicle interactions occur. While prior research has primarily focused on intersections and midblock locations, bus stops have often been treated as secondary contributors rather than as distinct sites requiring targeted safety assessments.   This has left a critical gap in understanding how traffic exposure, roadway characteristics, and bus stop design features specifically influence pedetrain crash risks around bus stop locations. To address these gaps, this study develops a comprehensive framework focused on pedestrian safety in the vicinity of bus stops. The proposed approach employs a Random Parameters Negative Binomial-Lindley (RPNB-L) model to account for unobserved heterogeneity and site-specific variability. Using data from 596 bus stops in Fort Worth, Texas (2018–2022), the model identifies that higher pedestrian crash frequencies are significantly associated with increased AADT, elevated boarding activity, and the absence of key safety elements such as crosswalks, medians, and lighting. Conversely, far-side bus stop placement, signalized intersections, sidewalks, and mixed-use development are associated with lower crash risks. Roads near schools and those with speed limits of $\leq$35 mph show elevated crash risk. To support proactive safety management, the study integrates a Full Bayes-based Potential for Safety Improvement (PSI) metric, enabling the identification of hazardous stops and high-risk corridors. By unifying advanced count-based modeling with strategic risk prioritization, this research offers actionable, data-driven insights for improving pedestrian safety near bus stops.

\end{abstract}


\maketitle

\section{Introduction}

\begin{figure}
\centering
\resizebox*{15cm}{!}{\includegraphics{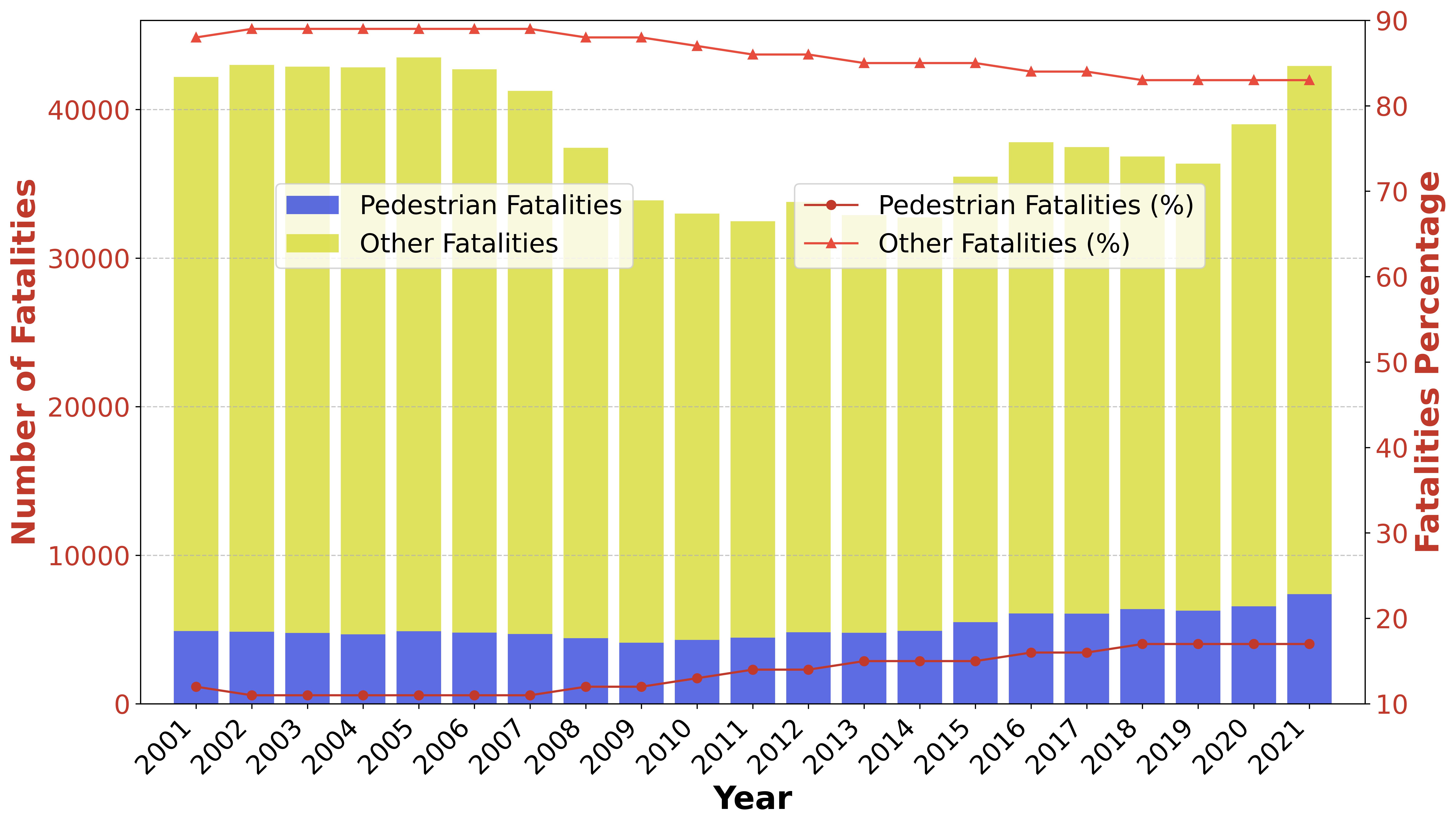}}
\caption{Pedestrian and vehicle fatalities (2001-2021)} \label{fig:1}
\end{figure}

Pedestrian safety in the United States has emerged as a critical public health issue, with pedestrian fatalities increasing substantially even as overall vehicle-related deaths have declined. Between 2020 and 2021, pedestrian fatalities increased by 12.54\%, and over the past two decades (2001–2021), pedestrian deaths increased by 50.74\%, while vehicle occupant fatalities declined by 4.67\% (\citep{stewart2023overview}). Since 2009, pedestrian fatalities have escalated by 59\%,  highlighting the need for enhanced safety measures (\citep{sanchez2024longitudinal}). In Texas, these trends are similarly concerning, with pedestrians accounting for 20\% of traffic-related deaths in 2020 (\citep{stewart2023overview}). In 2021 alone, 5370 pedestrian crashes resulted in 843 fatalities and 1467 severe injuries. Due to their vulnerability, pedestrians are approximately 1.5 times more likely to suffer fatal injuries compared to vehicle occupants (\citep{li2017analyzing, shinar2012safety, beck2007motor}). Fig. \ref{fig:1} illustrates the widening gap between pedestrian and vehicle fatalities, emphasizing the need for targeted safety interventions.

Research on pedestrian crashes has predominantly explored contributing factors and hazardous locations (\citep{xie2017analysis,lee2015multi}), demonstrating that roadway design, vehicle characteristics, and user behaviors collectively shape crash outcomes. Such findings support evidence-based strategies for enhancing pedestrian safety, improving infrastructure, and informing policy decisions (\citep{das2019supervised}). While intersections (\citep{haleem2015analyzing}) and mid-block areas (\citep{toran2017modelling, quistberg2015multilevel}) have received significant attention,  pedestrian crashes occurring specifically at or near bus stops have been comparatively neglected. Typically, bus stops have been viewed merely as contributing factors rather than distinct safety assessment sites, leaving a notable gap in understanding the unique risk factors and safety conditions associated with these locations. Factors such as traffic exposure, roadway environment, and specific bus stop designs have not been comprehensively evaluated in relation to pedestrian crashes, despite their potential to significantly influence crash risk. Additionally, the lack of a standardized analytical framework limits the identification and prioritization of hazardous bus stops within transit corridors.

Bus stops inherently present distinct safety challenges due to high pedestrian volumes, frequent street crossings by pedestrians rushing to catch buses, and complex traffic dynamics (\citep{yendra2024comparison}). Prior studies have indicated that the placement of bus stops near intersections or along arterial roads increases pedestrian crash risks significantly  (\citep{hu2018examination}). Campbell et al. (\citeyear{campbell2003review}) reported that approximately 2\% of urban pedestrian crashes occur in the vicinity of bus stops. Meanwhile, AASHTO (\citeyear{national2010highway}) has noted that having one or two bus stops within 1000 feet of an intersection raises pedestrian crash risks by 2.78 times, and by 4.15 times when more than two stops are present. Moreover, the growing prevalence of larger vehicles further exacerbates these risks (\citep{tyndall2021pedestrian}), particularly on busy transit corridors where multiple bus stops may be closely spaced (\citep{hess2004pedestrian}), making transit corridors a focal point of concern (\citep{craig2019pedestrian,ulak2021stop}).

Moreover, existing analytical frameworks often rely on traditional statistical models such as Poisson and Negative Binomial (NB), which frequently fail to accommodate the unique characteristics of pedestrian crash data, including randomness and a high proportion of zero-crash observations. These models assume that all locations have a probability of experiencing crashes. Although Zero-Inflated Poisson (ZIP) and Zero-Inflated Negative Binomial (ZINB) models attempt to address excess zeroes, these approaches often emphasize model fit over interpretability, potentially limiting their practical utility (\citep{lord2005poisson,lord2007further}). Alternative models, such as Negative Binomial-Lindley (NB-L) (\citep{geedipally2012negative, lord2011negative}) and and Negative Binomial-Generalized Exponential (NB-GE)(\citep{vangala2015exploring}), have demonstrated superior performance for highly skewed datasets.  Few studies compared the performance of ZINB, NB, NB-L, NB-GE (\citep{lord2011negative, vangala2015exploring}), these have shown that NB-L provides superior statistical fit, particularly for highly skewed datasets. Another persistent challenge lies in accounting for unobserved heterogeneity caused by spatial correlations, temporal variability, and omitted variables (\citep{mannering2016unobserved}). Random Parameters (RP) methods directly address this issue by allowing model coefficients to vary across sites, with the RPNB-L model showing particular promise when excess zeros and unobserved heterogeneity overlap (\citep{rusli2018applying}).

Accordingly, the main objective of this study is to develop a comprehensive framework for assessing pedestrian safety near bus stops, one that explicitly accounts for traffic exposure, roadway characteristics, and bus stop design features. More specifically, this study (i) proposes a pedestrian crash analysis framework tailored to bus stops along urban transit corridors; (ii) applies a random parameters approach to capture unobserved heterogeneity driven by site-specific variations; and (iii) employs a Full Bayes (FB) based Potential for Safety Improvement (PSI) metric to identify and prioritize high-risk transit corridors proactively. By addressing notable gaps in the existing literature, this work offers a systematic approach to understanding and mitigating pedestrian crashes near bus stops.

To accomplish these goals, the study integrates five years (2018–2022) of pedestrian crash data from 596 bus stop locations in Fort Worth, Texas, with complementary datasets on roadway conditions, traffic exposure, and bus stop design characteristics. Traffic exposure data were obtained from local agency records, while roadway and design attributes were extracted using the Rhino database and Google Earth \& Street View. To effectively capture site-specific variation and unobserved heterogeneity in crash frequencies, the study adopts the RPNB-L model as its core analytical method. Baseline models, including NB-L and RPNB-GE, are also estimated for comparison. Additionally, the Full Bayes-based PSI metric is applied to systematically identify hazardous bus stops and transit corridors, enabling a proactive, data-driven safety management approach. By bridging these critical gaps in the literature, this study contributes a novel, transferable framework for evaluating and improving pedestrian safety near bus stops in urban settings.

The structure of this study is organized as follows: Section \ref{sec2} provides an extensive review of the relevant literature. Section \ref{sec3} describes the methodology, model estimation process, inference, and validation. Section \ref{sec4} covers the data description. The model estimation results and a brief discussion are presented in Section \ref{sec5}. Section \ref{sec 6} summarizes the key points and presents the conclusions.

\section{Background \label{sec2}}

Bus stops are critical interaction points where pedestrians, cyclists, passengers, and vehicles frequently converge, increasing the likelihood of crashes (\citep{torbic2010pedestrian}). Vulnerable road users, such as pedestrians, passengers, etc., face significant safety risks near these locations (\citep{zhang2023revealing}). Studies indicate that 89\% of high-crash locations are within 150 feet of bus stops, and 90\% occurring within 70 feet of crosswalks (\citep{walgren1998using}). These findings highlight the need for enhanced pedestrian safety near bus stops. 

Various methods have been employed to analyze pedestrian safety, including historical crash data analysis (\citep{craig2019pedestrian, hess2004pedestrian}), geography information system-based safety inspections (\citep{yu2024impact}), road user observations (\citep{akintayo2022safety}), and traffic conflict techniques (\citep{zhang2023revealing}). Over time, crash-based analyses have utilized statistical models to establish relationships between crash frequency and explanatory variables (\citep{thakali2015identification}). These approaches provide a comprehensive understanding of influencing factors, such as exposure-related behaviors, roadway conditions, bus stop design elements (\citep{clifton2009severity,craig2019pedestrian,haleem2015analyzing,hu2018examination,zahabi2011estimating,zegeer2012pedestrian}). However, prior studies have often lacked a holistic approach, potentially leading to misleading conclusions about the effects of certain factors.

Pedestrian activity is particularly high near bus stops, especially in high-traffic areas such as schools and commercial zones, which heightens the risk of collisions (\citep{geedipally2021effects,craig2019pedestrian, hess2004pedestrian}). In addition to pedestrian-vehicle interactions, bus stops often serve as high-conflict areas for pedestrian-cyclist interactions, particularly where segregated bike paths or shared-use lanes exist. Afghari et al. (\citeyear{afghari2014pedestrian}) examined such conflicts in Montreal using automated video-based conflict analysis. Their study found that pedestrians frequently took evasive actions (e.g., stopping or running) to avoid cyclists, whereas cyclists generally maintained their speeds. Risky pedestrian behaviors, such as crossing midblock or rushing to board buses, further increase crash risks. Risky pedestrian behaviors, such as crossing without waiting for signals or rushing to board buses, further contribute to crash risks (\citep{quistberg2015bus}). Additionally, crossing roads at midblock instead of using designated crosswalks significantly increases crash potential (\citep{pessaro2017impact}). A study found that 59\% of pedestrian crashes occurred when individuals rushed to catch a bus (\citep{wretstrand2014safety}). Additionally, waiting in undesignated areas may cause buses to stop improperly, increasing the likelihood of crashes (\citep{chand2017improper}). These behaviors underscore passenger activity as a critical factor influencing crash frequency, with boarding and alighting volumes serving as key indicators of pedestrian risk near bus stops.

Bus stop sign infrastructure is essential to pedestrian safety. Clear and visible signage ensures that pedestrians wait in designated areas and helps bus drivers identify stopping points. When signage is obstructed by trees or advertisements, the area becomes more hazardous (\citep{rossetti2020field}). Bus shelters provide accessible waiting spaces, and the absence of such facilities has been linked to the most hazardous bus stop locations (\cite{salum2024toolkit}). These risks are further exacerbated by inadequate pedestrian infrastructure, such as marked crosswalks or pedestrian signals (\citep{pulugurtha2008hazardous}). Several studies underscore the importance of lighting for safety at bus stops, especially at nighttime (\citep{rossetti2020field,mukherjee2023built}), with poor lighting linked to higher pedestrian fatality rates, especially in sparsely populated areas (\citep{lakhotia2020pedestrian}). Many studies recommend improved lighting at bus stops and their approaches to enhance safety (\citep{salum2024toolkit,pessaro2017impact}).

The location of bus stops also influences pedestrian crash risks due to factors such as driver distraction and poor yielding behavior (\citep{craig2019pedestrian}). Fitzpatrick and Nowlin (\citeyear{fitzpatrick1997effects}) examined the safety impacts of bus stop placement and found that far-side stops (positioned after an intersection) experience fewer crashes than near-side stops due to reduced conflicts with right-turning vehicles. However, far-side stops can obstruct intersections during peak hours, diminishing visibility for both drivers and pedestrians  (\citep{texas1996guidelines}). 

Safe and accessible pedestrian infrastructure around bus stops strongly correlates with increased transit access and safety (\citep{sukor2020safety}). The absence of sidewalks forces pedestrians to walk on roadways, increasing their exposure to traffic and crash risk (\citep{rossetti2020field}). Similarly, the presence of crosswalks improves pedestrian safety (\citep{ulak2021stop}), yet over 30\% of bus stops lack them, making these areas more hazardous (\citep{tiboni2013implementing}). Studies indicate that a substantial proportion of pedestrian crashes occur due to non-use of crosswalks, with two-thirds of pedestrian fatalities occurring under such conditions (\citep{clifton2009severity}). At unsignalized intersections, pedestrian injury severity is significantly influenced by crosswalk availability (\citep{clifton2009severity}). Additionally, bus stops near signalized intersections exhibit higher pedestrian crash rates due to increased pedestrian volume and transit activity (\citep{pulugurtha2011pedestrian}). Several studies acknowledge this association, emphasizing that bus stops near intersections pose heightened pedestrian risks (\citep{harwood2008pedestrian,kucskapan2022pedestrian,walgren1998using}). Zeeger \& Bushell (\citeyear{zegeer2012pedestrian}) found that the likelihood of pedestrian crashes increases significantly when bus stops are located within 1,000 feet of an intersection, especially when combined with high traffic and pedestrian volumes, multiple lanes without a median, and nearby schools. Similarly, Torbic et al. (\citeyear{torbic2010pedestrian}) developed a method to predict vehicle-pedestrian crashes, showing that crashes are four times more likely when a bus stop is within 1,000 feet of a signalized intersection, mainly when schools or alcohol retail stores are present.

Several studies have examined the relationship between the built environment, traffic characteristics, and crash frequency, considering factors such as traffic volume, speed, and roadway geometry (\citep{hess2004pedestrian, ulak2021stop, lakhotia2020pedestrian, tiboni2013implementing}). For example, high average annual daily traffic (AADT) near bus stops increases potential conflicts between vehicles and pedestrians. High traffic speeds are also a major safety risk, particularly in urban areas where pedestrian activity is high (\citep{rossetti2020field}). Furthermore, bus stops located along high-speed roads are linked to increased pedestrian fatalities (\citep{hess2004pedestrian, lakhotia2020pedestrian, pulugurtha2008hazardous}). Roadway width also plays a role in pedestrian risk, as wider roads complicate pedestrian crossings and increase crash likelihood, whereas narrower roads may result in more pedestrian-vehicle conflicts  (\citep{zhang2023revealing}). The presence of a median near bus stops enhances safety by providing pedestrians with a refuge area, allowing them to cross in stages  (\citep{pessaro2017impact, yu2024impact}). Land use also plays a crucial role in pedestrian safety near bus stops. High-activity areas, such as shopping centers, schools, and workplaces, attract significant pedestrian and vehicular traffic, increasing collision risks  (\citep{hess2004pedestrian}). Areas with high vehicular activity, mainly commercial zones like big box stores, are associated with more pedestrian crashes due to increased traffic flow and parking-related movements (\citep{yu2024impact}).

Several researchers have conducted investigations about identifying hotspots with a high incidence of pedestrian-vehicle collisions and bus stops that pose safety risks (\citep{truong2011using,ulak2021stop}).  Truong \& Somenahalli (\citeyear{truong2011using}) proposed a GIS-based approach utilizing spatial autocorrelation analysis and severity indices to rank unsafe bus stops based on pedestrian crash data. Ulak et al. (\citeyear{ulak2021stop}) introduced the Bus Stop Safety Index (SSI), a quantitative measure that incorporates injury severities, pedestrian crash proximity, socio-demographic factors, traffic conditions, and transit stop attributes to identify high-risk bus stops in Palm Beach County, FL. This metric facilitates targeted interventions to enhance pedestrian safety and improve urban transit system accessibility.

While previous studies have acknowledged the significance of bus stops in pedestrian safety, they often fail to provide a comprehensive framework that integrates exposure characteristics, roadway environments, and bus stop design elements in assessing crash risk. Most studies consider bus stops as secondary variables rather than directly examining how infrastructure, traffic, and environmental factors interact to influence the frequency of crashes. To address this gap, the present study employs advanced statistical models that integrate diverse categorical variables to quantify crash risks and identify hazardous transit corridors. These models aim to generate actionable insights, supporting transportation agencies in developing targeted interventions to enhance pedestrian safety near bus stops.

\section{Methodology \label{sec3}}

The primary objective of this study is to develop and apply a robust statistical modeling framework to investigate the relationship between pedestrian crashes and key contributing factors near bus stops. Based on the research gaps and objectives outlined in the previous sections, multiple statistical modeling approaches have been employed, informed by established crash frequency modeling research. This section provides a concise overview of the selected modeling methodologies and describes the evaluation criteria used to compare their performance and interpretability.

\subsection{Negative Binomial-Lindley (NB-L)\label{sec3.1}}

The NB-L model (\cite{geedipally2012negative, rahman2016use}) is an extension of the standard NB model designed to handle highly overdispersed count data. Geedipally et al. (\citeyear{geedipally2012negative}) demonstrated that the NB-L model performs better than the NB model when the data contains many zeros or is characterized by a long tail. The NB-L model introduces an additional layer of flexibility by combining the NB distribution with the Lindley distribution to better account for unobserved heterogeneity and excess variation in the data. Therefore, the mixture of the NB and Lindley distributions is as follows:

\begin{equation}
P(Y_i = y_i | \mu_i, \phi, \theta) = \int \text{NB}(y_i | \phi, \lambda_i \mu_i)  \text{Lindley}(\lambda_i | \theta) \, d\lambda_i
\label{Equation 5}
\end{equation}

The random error term is represented by \(\lambda_i\), while the Lindley distribution is denoted by \(\theta\). In Equation (\ref{Equation 5}), both \(\lambda_i\) and \(Y_i\) are assumed to follow a NB distribution, with the mean defined as \(\lambda_i \mu_i\). The corresponding inverse dispersion parameter is expressed as \(\phi = \frac{1}{\alpha}\), where \(\lambda_i\) is governed by the Lindley distribution parameter \(\theta\).

\bigskip

Here, the Lindley distribution can be presented as (\citep{lord2021highway}) is defined as:

\begin{equation}
\text{Lindley} (x | \theta) = \frac{\theta^2}{\theta + 1} (1 + x) e^{-\theta x}, \quad \theta > 0, x > 0
\label{Equation 6}
\end{equation}

The mean response function for the NB-L can be expressed as:

\begin{equation}
 E(Y_i = y_i) = \mu_i \cdot E(\lambda_i)
\label{Equation 7}
\end{equation}

Where $\mu_i$=$\exp(b_0 + \sum_{j=1}^{q} b_j X_j)$  represents the expected number of crashes, and $E(\lambda_i)$ = $\frac{\theta + 2}{\theta(\theta + 1)}$ is the expected value of the Lindley-distributed term. Replacing the values for $\mu_i$ and $E(\lambda_i)$, the mean response function becomes:

\begin{equation}
E(Y_i = y_i) = \exp\left(b_0 + \sum_{j=1}^{k} b_j X_j\right) \cdot \frac{\theta + 2}{\theta(\theta + 1)}
= \exp\left(b_0 + \log\left(\frac{\theta + 2}{\theta(\theta + 1)}\right) + \sum_{j=1}^{k} b_j X_j\right)
\label{Equation 8}
\end{equation}

To simplify, we incorporate the term $\log\left(\frac{\theta + 2}{\theta(\theta + 1)}\right)$ into a new intercept term, $b'_0$, yielding:

\begin{equation}
E(Y_i = y_i) = \exp\left(b'_0 + \sum_{j=1}^{k} b_j X_j\right)
\label{Equation 9}
\end{equation}

Where:

\[
b'_0 = b_0 + \log\left(\frac{\theta + 2}{\theta(\theta + 1)}\right)
\]

Thus, the NB-L mean response function accounts for over-dispersion and excess zeros by adjusting the intercept term to include the effects of the Lindley distribution.

\bigskip

The Lindley distribution is a mixture of two Gamma distributions as follows (\citep{zamani2010negative}):

\begin{equation}
\lambda_i \sim \frac{1}{1 + \theta} \text{Gamma}(2, \theta) + \frac{\theta}{1 + \theta} \text{Gamma}(1, \theta)
\label{Equation 10}
\end{equation}

The Equation \ref{Equation 10} alternatively, it can be restructured as according to \citep{lord2021highway}

\begin{equation}
\lambda_i \sim \sum \left\{ \text{Gamma}(1+z_i, \theta) 
 \text{Bernoulli}\left(z_i; \frac{1}{1 + \theta}\right) \right\}
\label{Equation 11}
\end{equation}

Therefore, the NB-L model can be written as the following hierarchical model (\citep{geedipally2012negative, lord2021highway, khodadadi2023evaluating}):

\begin{equation}
P(Y_i = y_i; \phi, \mu_i | \lambda_i) = \text{NB}(y_i; \phi, \lambda_i \mu_i)
\label{Equation 12}
\end{equation}

\begin{alignat}{2}
\mu_i &= \exp\left(b_0 + \sum_{j=1}^{q} b_j X_j\right) \\
\lambda_i &\sim \text{Gamma}(1 + z_i, \theta) \\
z_i &\sim \text{Bernoulli}\left(\frac{1}{1 + \theta}\right)
\end{alignat}

Prior distributions must be specified to estimate the model parameters using a Bayesian framework. These priors reflect past knowledge or assumptions about the parameters. The random error $\lambda$ term is given a non-informative prior from a gamma distribution, while the gamma shape parameter is governed by a Bernoulli distribution with probability \(1/(1 + \theta)\). When weakly informative priors are used, the Lindley distribution can influence the model more than the NB component. One challenge in Bayesian estimation is the poor mixing of the MCMC process caused by the intercept and error term correlation. Geedipally et al.( \citeyear{geedipally2012negative}) recommend employing informative priors when possible to mitigate this issue. A prior should be chosen to ensure that \(E(\varepsilon) = 1\) limits the influence of the Lindley distribution. Geedipally et al. (\citeyear{geedipally2012negative}) advocate for the use of a prior on \(1/(1 + \theta)\) that follows a beta distribution, specifically Beta(n/3, n/2), where \(n\) represents the number of observations. This approach enhances model stability and improves the accuracy of parameter estimation. 

\subsection{Random Parameters Negative Binomial-Lindley (RPNB-L)\label{sec3.2}}

Random parameters (RP) models an alternative way to account for unobserved heterogeneity by allowing parameters to vary from one observation to another (\cite{anastasopoulos2009note,barua2016multivariate,shaon2018developing, rusli2018applying, khan2023effects}). To form this randomized variable, lets assume the coefficient $\beta_{ij}$ associated with the $j$-th covariate for the $i$-th site and which can be expressed as:

\begin{equation}
b_{ij} = b_j + v_{ij}
\label{Equation 13}
\end{equation}

Where $b_j$ represents the fixed mean estimate of the parameter, and $v_{ij}$ is the random term, which typically follows a normal distribution with a mean of zero and a variance of $\sigma^2$. This $v_{ij}$ is included in the model if its standard deviation is significantly different from zero; otherwise, a fixed parameter $b_j$ is applied uniformly across all observations (\citep{el2009urban, anastasopoulos2009note}). The probability mass function (pmf) of the RPNB model is given by:

As mentioned above, the NB-L model itself is the RP model since it accommodates the Lindley distribution by introducing site-specific effects, which coefficient fully transform to randomized by introducing RPNB-L, meaning that the coefficients of the covariates vary across sites, not just the Lindley portion. This leads to a hierarchical model structure, according to Shaon et al.(\citeyear{shaon2018developing}):

\begin{align}
P(Y_i = y_i; \phi, \mu_i \mid \lambda_i) &= \text{NB}(y_i; \phi, \lambda_i \mu_i) \nonumber \\
\mu_i &= \exp\left(b_0 + \sum_{j=1}^{k} b_{ij} X_{ij}\right)
\label{Equation 14}
\end{align}

In this hierarchical structure, the random effect $\lambda_i$ follows a gamma distribution, $\lambda_i \sim (\lambda|1 + z_i, \theta)$, and $z_i$ follows a Bernoulli distribution $z_i \sim \text{Bernoulli}\left(z|\frac{1}{1 + \theta}\right)$. The random parameter $b_{ij} = b_j + v_{ij}$ captures site-specific variability, where $v_{ij} \sim \mathcal{N}(0, \sigma_j^2)$.

\bigskip

A known issue with RPNB-L models, especially in a Bayesian framework, is poor mixing of the MCMC chains, mainly due to correlations between the intercept and coefficients, both of which vary across observations. One solution is to standardize the coefficients before using them in the model. After the MCMC chains converge, the standardized coefficients can be transformed back to their original scale using the following transformations for a more straightforward interpretation:

\begin{equation}
x_{ij}^* = \frac{x_{ij} - \mu_j}{\sigma_j}
\label{Equation 15}
\end{equation}

where $\mu_j$ and $\sigma_j$ are the mean and standard deviation of the $j$-th covariate. After the analysis, the estimated coefficients are transformed back to the original scale using (\citep{gelfand1995efficient}):

\begin{equation}
b_1 = \frac{b_1^*}{\sigma_1}, \quad \dots, \quad b_q = \frac{b_q^*}{\sigma_q}, \quad b_0 = b_0^* - \sum_{j=1}^{q} \frac{b_j \mu_j}{\sigma_j}
\label{Equation 16}
\end{equation}

The random parameters $b_{ij}$ are modeled using a normal prior:

\begin{equation}
b_{ij} \sim \mathcal{N}(b_j, \sigma_j^2), \quad \frac{1}{\sigma_j^2} \sim \Gamma(0.01, 0.01)
\label{Equation 17}
\end{equation}

Previous research has shown that normal distributions typically provide the best fit for modeling the RP compared (\citep{anastasopoulos2009note}) to using lognormal, uniform, gamma, triangular, etc., which helps ensure adequate MCMC mixing and improves the robustness of the model's estimates.

\subsection{Random Parameters Negative Binomial – Generalized Exponential (RPNB-GE)\label{sec3.3}}

The RPNB-GE is an extension of the standard NB model, formulated to account for overdispersion, excess zeros, and unobserved heterogeneity in crash history data (\cite{vangala2015exploring}). The GE distribution is used to introduce an additional layer of flexibility for capturing these variations. For a deeper understanding, readers are directed to (\citep{aryuyuen2013negative}). The NB-GE model assumes that the mean crash rate is adjusted by a Generalized Exponential-distributed random effect, leading to the following formulation:

\begin{equation}
P(Y_i = y_i | \mu_i, \phi, a, b) = \int \text{NB}(y_i | \phi, \theta_i) \text{GE}(\delta_i | a, b) \, d\delta_i
\label{Equation 18}
\end{equation}

Where:
\begin{align*}
\theta_i &= \mu_i \delta_i \\
\delta_i &\sim \text{Generalized Exponential} (a, b)
\end{align*}

Here, $\theta_i$ is the product of $\mu_i$ and $\delta_i$, representing the mean crash rate adjusted by the Generalized Exponential-distributed term $\delta_i$. The random effect $\delta_i$ distribution follows as  $ Generalized Exponential (a, b)$, where $a$ and $b$ are the distribution parameters. 

\bigskip

\subsection{Model estimation \label{sec3.4}}
The previous sections outlined the posterior estimation framework for all models, which was implemented using a Bayesian approach through Markov Chain Monte Carlo (MCMC) simulation in OpenBUGS. Given the bias and auto-correlation inherent in MCMC samples, this study ran numerous iterations to mitigate these effects. Multiple simulation chains were executed to ensure robust convergence; each initialized with different starting values. Specifically, three chains were run for 80,000 iterations per parameter, with the first 30,000 samples discarded as burn-in to eliminate potential bias. The remaining 50,000 iterations were used to estimate posterior distributions. Convergence was evaluated by visually inspecting trace plots, which show the progression of chains from various starting points, and by calculating the Brooks-Gelman-Rubin (BGR) statistic for each parameter (\cite{el2009urban}). A BGR value stabilized near 1 (below 1.1) indicated effective convergence (\cite{gelman1992inference,mitra2007nature}). Additionally, each parameter's Monte Carlo (MC) error was kept below 3\% of the posterior standard deviation to ensure reliable parameter estimates.

The dependent variable considered pedestrian-vehicle crashes with all severity levels (KABCO) included. The study highlights that only explanatory variables with a 5\% significance level were included in the results, indicating the use of a 95\% CI to identify key crash-contributing factors. In Bayesian modeling, the random parameter framework introduces an additional layer of complexity by allowing parameters to vary across observations, capturing unobserved heterogeneity. However, when the variability in these parameters is minimal, their inclusion as random parameters can unnecessarily complicate the model without adding meaningful value. To balance complexity and performance, carefully determining which parameters benefit from this hierarchical structure is crucial.

\subsection{Model performance\label{sec3.5}}

The Deviance Information Criterion (DIC) is widely used within the Bayesian framework for model selection, given the availability of numerous models. The principle of DIC is parsimony, aiming to find the simplest model that explains the most variation in the data. The DIC is calculated as:

\begin{equation}
DIC = \overline{D}(\Theta) + P_D \quad
\left[
\overline{D}(\Theta) = E[-2 \log L], \quad
P_D = \overline{D}(\Theta) - D(\overline{\Theta})
\right]
\label{eq18}
\end{equation}

Where \( \overline{D}(\Theta)\) is the posterior mean deviance indicating model fit, \( \Theta \) is the total number of parameters, \( P_D \) is the adequate number of parameters, L is the likelihood. Generally, a model with a lower DIC is preferred. A difference greater than 10 in DIC values between models strongly favors the model with the lower DIC. Differences between 5 and 10 are considered substantial, while differences less than 5 suggest that the models are competitive.
\bigskip

The evaluation process relied on metrics, Mean Absolute Error (MAE), and Root Mean Square Error (RMSE) to assess the performance of the models on the test set and identify the best combination of hyperparameters for each model. By comparing the performance of all models, we can locate the superior results in capturing the relationship between pedestrian crashes and risk factors (\cite{lord2021highway}). The MAE is calculated as:

\begin{equation}
\text{MAE} = \frac{1}{n} \sum \left| y_{\text{observed}} - y_{\text{predicted}} \right|
\label{eq19}
\end{equation}

Where $n$ represents the number of observations, and $y_{\text{observed}}$ and $y_{\text{predicted}}$ correspond to the observed and predicted crash counts, respectively. The RMSE is calculated as:

\begin{equation}
\text{RMSE} = \sqrt{\frac{1}{n} \sum \left( y_{\text{observed}} - y_{\text{predicted}} \right)^2}
\label{eq20}
\end{equation}

These metrics are commonly employed to assess the accuracy of the model predictions by comparing the observed and predicted values of crash counts. The MAE provides the average magnitude of the errors in a model's predictions. At the same time, the RMSE gives greater weight to more significant errors, offering a more sensitive measure of model performance.

\subsection{Marginal effects\label{sec3.6}}
Usually, the marginal effect (ME) is computed to explore a deeper understanding of the effects of explanatory variables (significant) on crashes. Which determines the unit change in expected crash frequency due to a unit changing explanatory variable \( X_j \), while all other explanatory variables must be constant. For explanatory variables whose values range from 0 to 1 (\citep{williams2012using}), calculating ME according to \cite{washington2020statistical}:

\begin{equation}
\text{ME}_{X_{ij}} = \frac{\partial \lambda_i}{\partial X_{ij}}
\end{equation}

Here, \( \text{ME}_{X_{ij}} \) is defined as the marginal effect of the $j^{th}$ explanatory variable at the $i^{th}$ observation, \(\lambda_i\) denotes the expected mean crash frequency, and \( X_j \) is the $j^{th}$ explanatory variable of interest.

\section{Data description \label{sec4}}

This section outlines the data foundation of the study. Subsection~\ref{sec4.1} describes the five years of pedestrian crash data obtained from TxDOT’s Crash Records Information System (CRIS) database. Subsection~\ref{sec4.2} details the complementary data on bus stop design, roadway features, and traffic exposure. Together, these datasets enable a comprehensive and context-rich assessment of pedestrian safety near bus stops.

\subsection{Pedestrian crash data\label{sec4.1}}

This study utilized bus stop data obtained from Trinity Metro, the public transit agency serving the Fort Worth metropolitan area. The data were accessed through the agency’s official website (\url{https://www.trinitymetro.org}) and included Geographic Information Systems (GIS) shapefiles containing spatial and attribute information for approximately 2500 bus stops. Each stop was identified by its corresponding main street and nearby cross-street and included details on amenities and wheelchair accessibility. In addition to the spatial data, Trinity Metro provided two supplementary spreadsheets reporting passenger boarding and alighting counts, disaggregated by stop location and service day type (weekday, Saturday, and Sunday). In the absence of direct pedestrian activity data, these counts served as a proxy for estimating relative crowding levels at each stop.

In alignment with the study’s objective, historical pedestrian crash data were systematically integrated with the stop dataset. The crash records for 2018 to 2022 were obtained from CRIS. This dataset included only reportable crashes, defined as incidents involving a fatality, injury, or property damage exceeding \$1000, and restricted to public roadways. To spatially associate pedestrian crashes with bus stop locations, a 250 ft buffer was established around each stop. This distance was informed by prior literature (e.g., \cite{craig2019pedestrian}), which noted that the majority of pedestrian crashes related to transit access occur within 150 feet of a stop. Building on this, and through a trial-and-error procedure aimed at generating reliable vehicle-pedestrian crash concentration maps, a 250 ft buffer was determined to provide an optimal balance. The threshold was sufficiently wide to capture pedestrian movements and crossing behaviors associated with bus stop access, yet narrow enough to reduce the likelihood of including unrelated incidents on adjacent roadways. To maintain spatial precision, crashes located within the buffer but occurring on cross-streets were manually reviewed and excluded from the analysis. Given the challenges in collecting detailed geometric and operational data for all bus stops, a targeted sampling strategy was adopted to enable a comprehensive yet manageable evaluation. Stops were selected using two clearly defined criteria to represent both high and low-risk environments:

\begin{enumerate}
\item Bus stops with at least one recorded pedestrian crash.
\item Bus stops exhibiting reasonable pedestrian activity (as indicated by boarding and alighting data) but with no recorded crashes.
\end{enumerate}

\begin{figure}[h]
    \centering
    \setlength{\abovecaptionskip}{0pt}
    \includegraphics[width=0.8\textwidth]{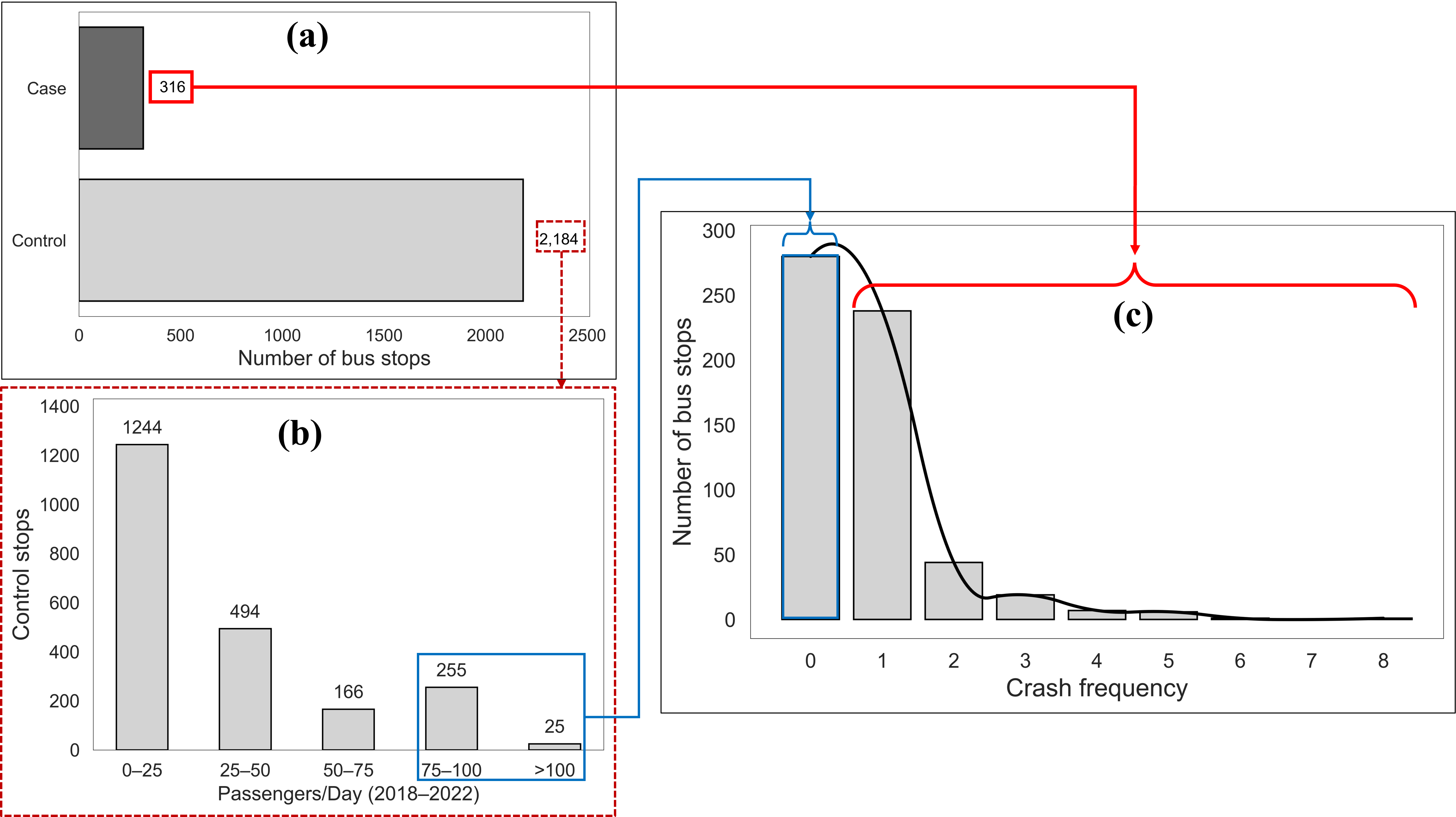}
    \caption{Distribution of "Case" and "Control" stops: (a) stops classification, (b) daily ridership at "Control" stops, (c) frequency distribution among selected stops.}
    \label{Pic:2}
\end{figure}

The first criterion captures stops with documented pedestrian crash history during the study period, enabling a reactive analysis of contributing factors. To ensure inclusivity across the full spectrum of crash severities, 316 bus stops with one or more pedestrian crashes were designated as "case" stops. Among these, 75.3\% had exactly one crash, 13.9\% had two crashes, and a small number had three or more, with a maximum of eight at a single stop. This distribution, shown in Fig.~\ref{Pic:2}(a \& c), follows a long-tail pattern, most crash-involved stops experience low crash frequencies. Applying a stricter threshold (e.g., two or more crashes) would have excluded nearly 90\% of these locations, greatly limiting sample diversity and statistical power. Including all stops with at least one crash thus ensures a more inclusive and representative crash risk analysis.

For the second criterion, a quantitative threshold was applied to define “reasonable pedestrian activity.” As illustrated in Fig.~\ref{Pic:2}(b), most non-crash (control) stops had low boarding and alighting volumes: 1244 stops had 0–25 passengers/day, and 494 had 25–50 passengers/day. Including these would have introduced a major imbalance in exposure relative to the 316 case stops. Although higher-activity categories (50–75 and $\geq$100 passengers/day) contained fewer stops (n = 166 and 25, respectively), they still offered limited comparability. To mitigate this imbalance and adhere to resource constraints during data collection and fieldwork, the study limited control stops to those with $\geq$75 daily passengers. This threshold captured the upper end of the exposure distribution and provided a control sample reasonably aligned in size and exposure level with the case group. Although alternative thresholds (e.g., $\geq$50 or $\geq$100 passengers/day) were not tested in this study, future research could assess the sensitivity of results to these variations. The adopted threshold in this study reflects a practical balance between analytical rigor and feasibility, ensuring a matched, exposure-informed comparison between crash-prone and non-crash locations. By incorporating both types of stops, the analysis identifies overrepresented risk factors at hazardous sites and potential protective features at safer ones, strengthening the overall understanding of pedestrian safety dynamics near bus stops.

\begin{figure}[h]
    \centering
    \setlength{\abovecaptionskip}{0pt}
    \includegraphics[width=0.8\textwidth]{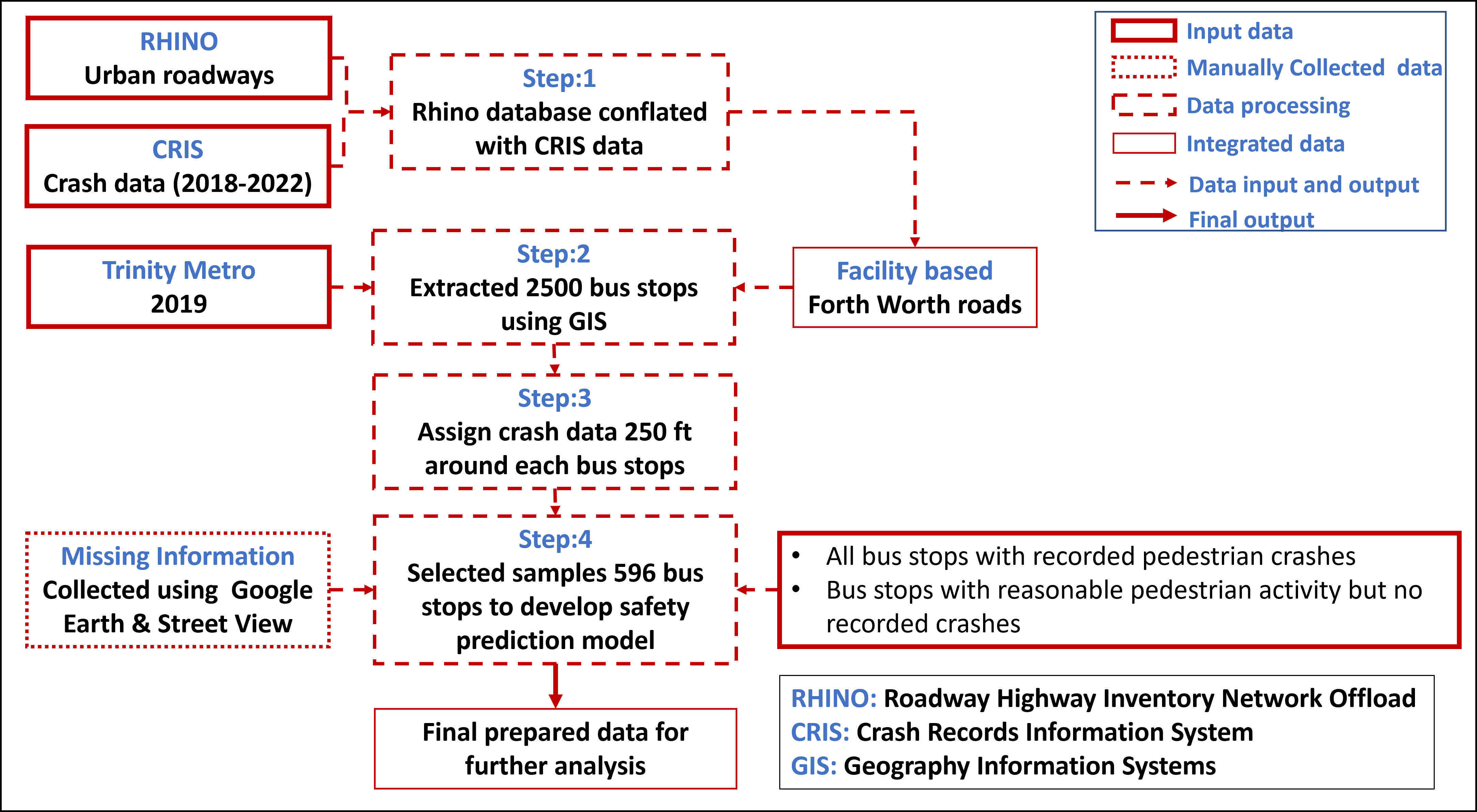}
    \caption{Flowchart for data preparation}
    \label{fig:2}
\end{figure}

\begin{figure}[h]
    \centering
    \setlength{\abovecaptionskip}{0pt}
    \includegraphics[width=0.7\textwidth]{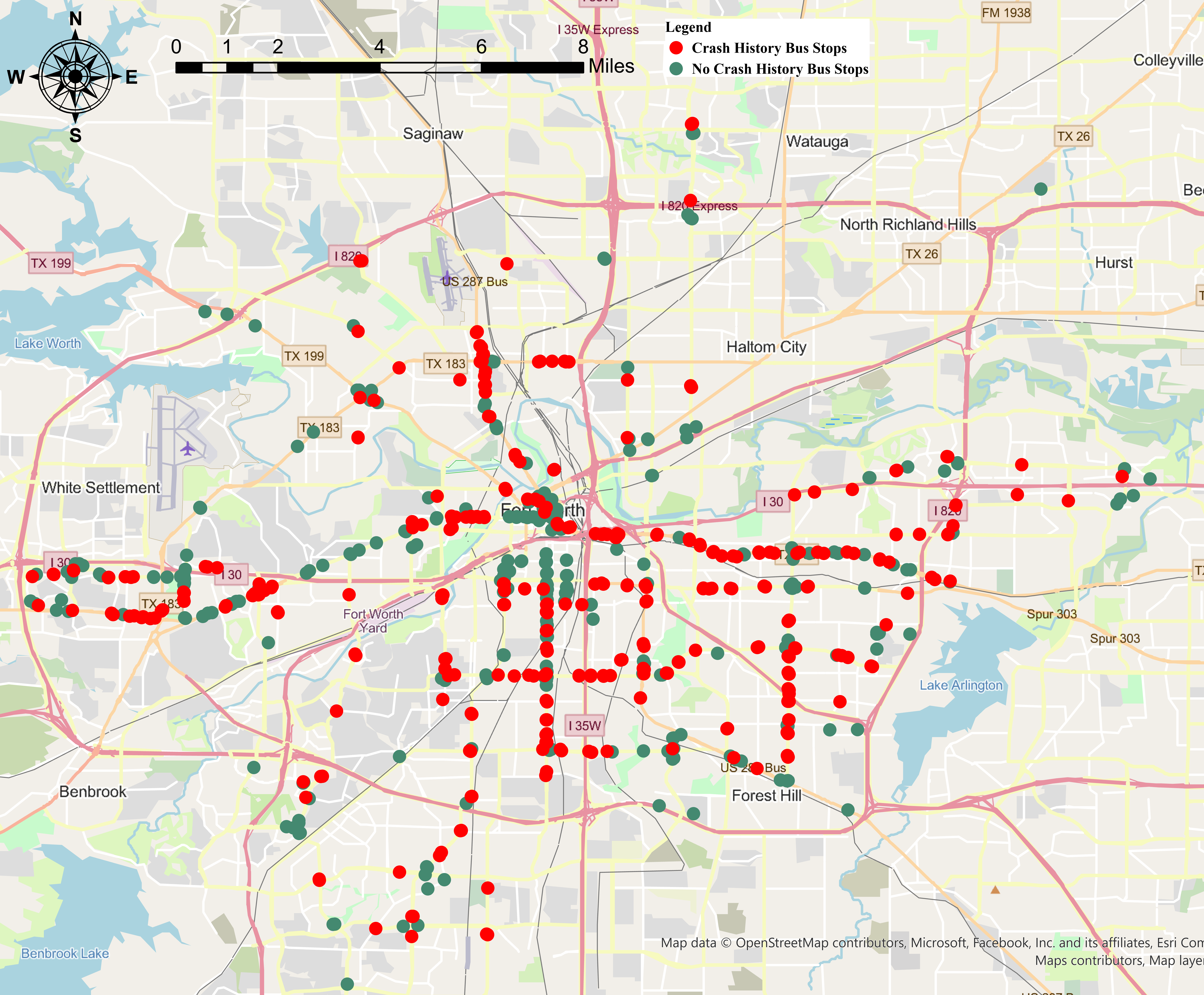}
    \caption{Sample bus stop locations for data collection in Fort Worth, Texas}
    \label{fig:3}
\end{figure}

Fig. \ref{fig:2} presents a flowchart summarizing the data preparation process, including the integration of pedestrian crash records with bus stop and roadway information. Based on the sampling criteria described earlier, a total of 596 bus stops were selected for detailed analysis (Fig.~\ref{fig:3}). To support this analysis, pedestrian crash data were integrated with roadway geometric and traffic characteristics extracted from TxDOT’s 2019 Roadway Highway Inventory Network Offload (RHiNo) database, a comprehensive GIS-based dataset.

The RHiNo database contains location-specific roadway information for both state and local roads, including system type, facility type, number of lanes, surface width, shoulder width, median type, and traffic volumes (current and historical). When key geometric features, such as the presence of the sidewalk, the availability of the crosswalk, or the lighting conditions, were missing, additional data were collected through Google Earth® and Google Street View®, which served as effective tools for field validation and remote assessment.

\begin{figure}[h]
    \centering
    \setlength{\abovecaptionskip}{0pt}
    \includegraphics[width=0.9\textwidth]{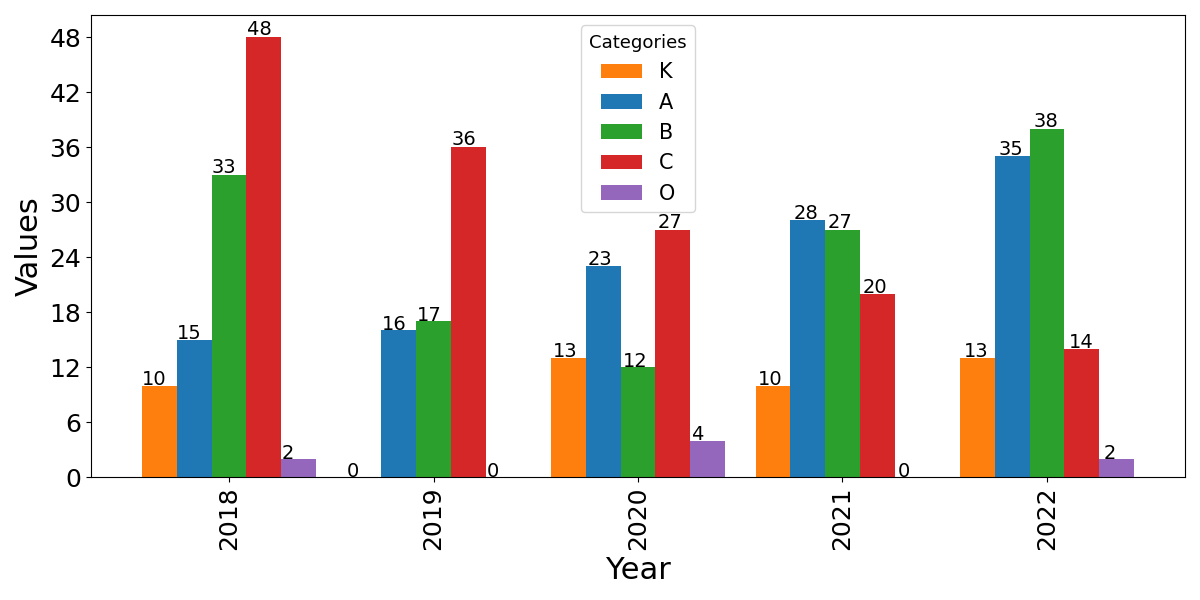}
    \caption{Yearly different severity levels of pedestrian crashes}
    \label{fig:4}
\end{figure}

\begin{figure}[h!]
    \centering
    \setlength{\abovecaptionskip}{0pt}
    \subcaptionbox{}
    {\includegraphics[width=0.30\textwidth]{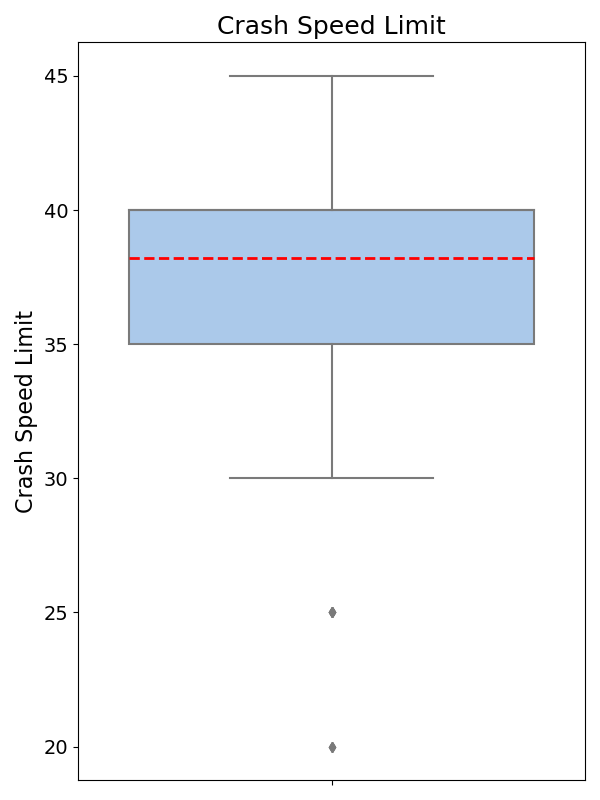}}
    \subcaptionbox{}{\includegraphics[width=0.30\textwidth]{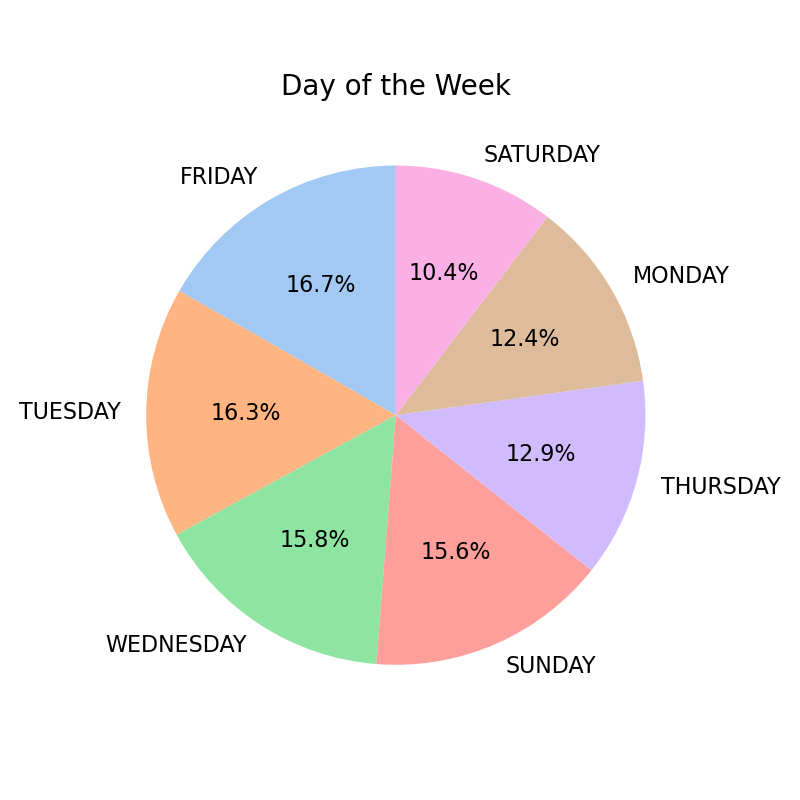}}
    \subcaptionbox{}{\includegraphics[width=0.35\textwidth]{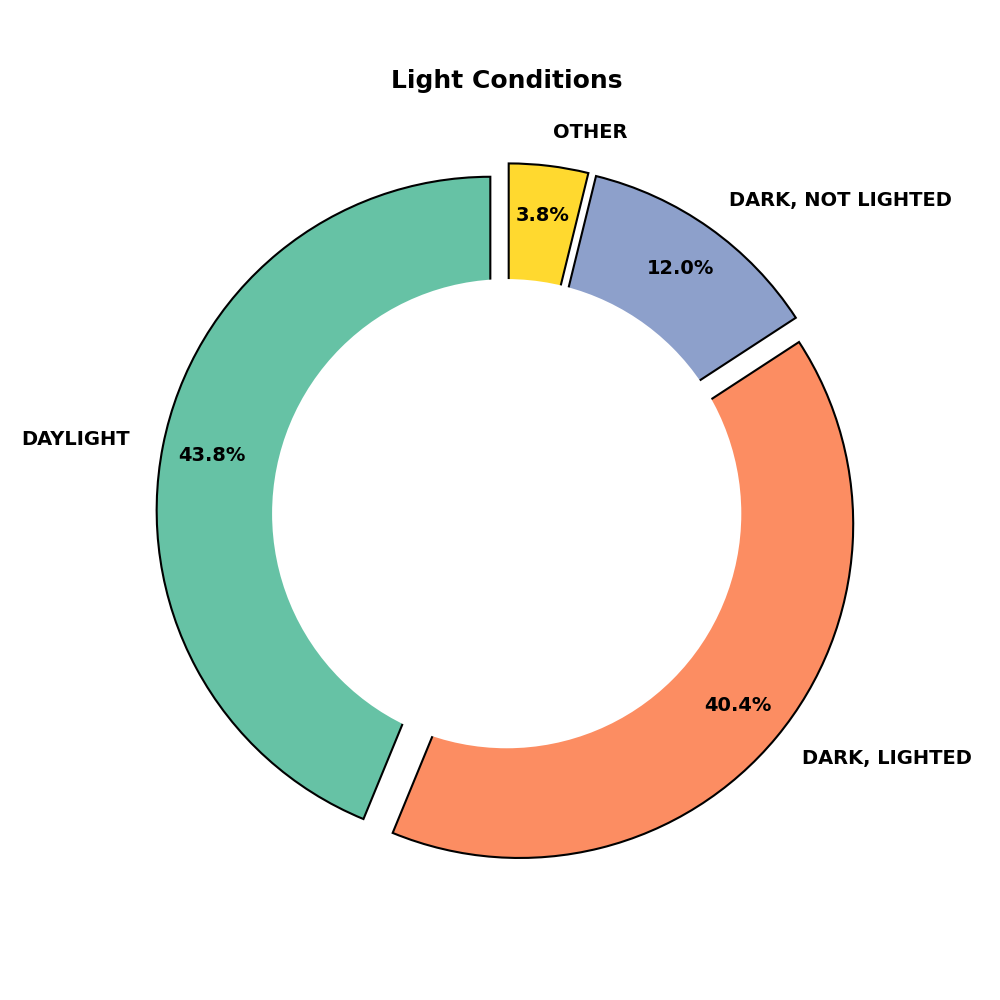}}
    \subcaptionbox{}{\includegraphics[width=0.35\textwidth]{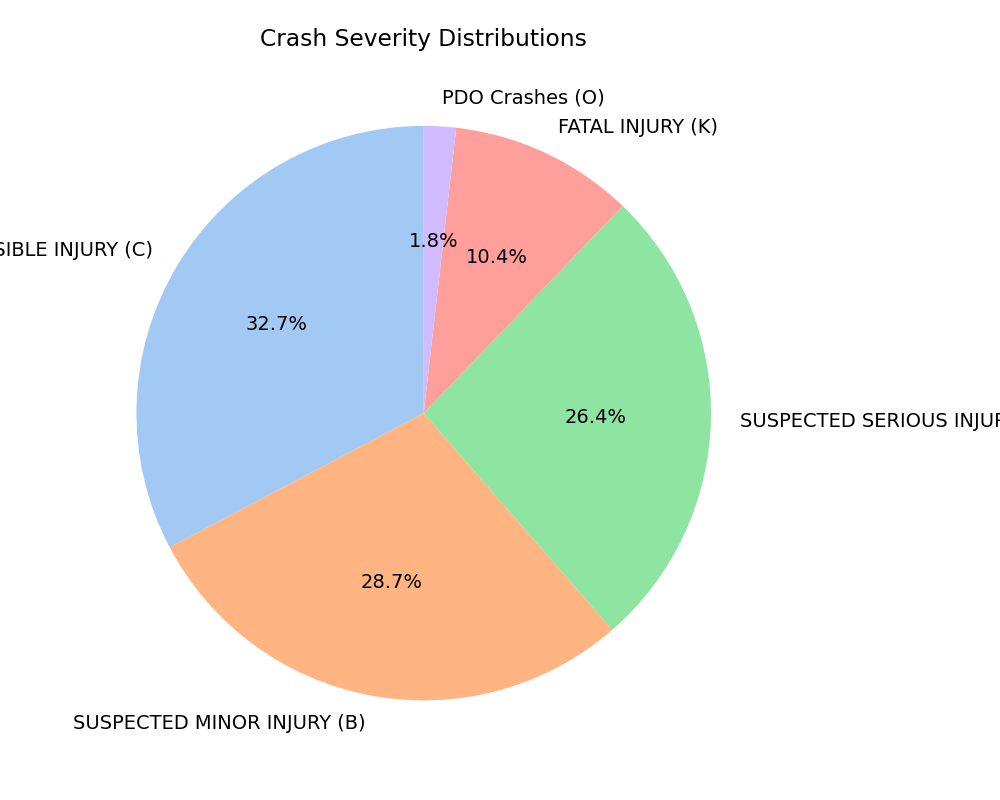}}
    \subcaptionbox{}{\includegraphics[width=0.35\textwidth]{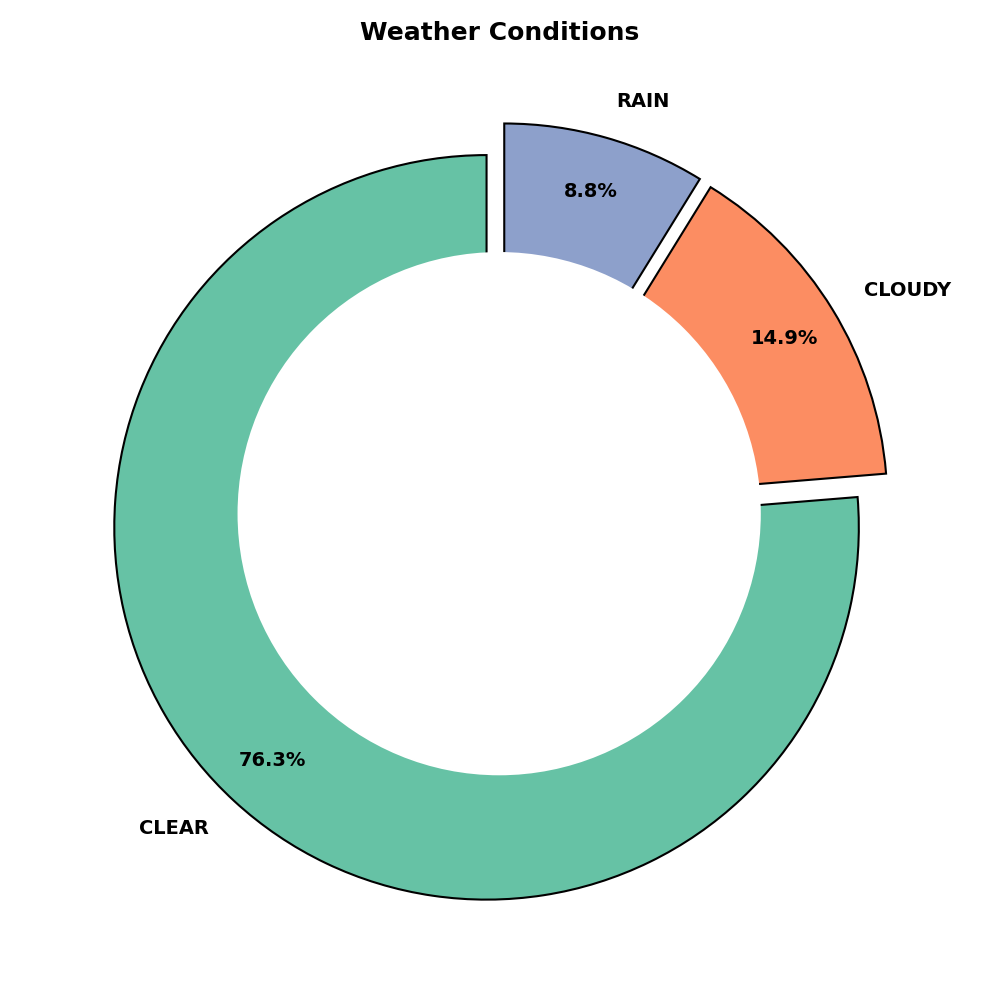}}
    \caption{Crash distribution characteristics}
    \label{fig:5}
\end{figure}

The final dataset included 448 pedestrian crashes, categorized by severity. Among these, 46 (10.3\%) were fatal (K-level), 394 (87.\%) involved injuries (A, B, or C-level), and 8 crashes (1.8\%) were classified as property damage only (PDO). The yearly distribution of crash severity is shown in Fig.~\ref{fig:4}, while Fig.~\ref{fig:5} illustrates general crash characteristics, including speed limit, day of the week, lighting, severity distribution, and weather conditions.

These underscore the substantial risk to pedestrians near bus stops—with nearly 98\% of crashes resulting in injuries or fatalities, the data highlight an urgent need for targeted safety improvements. The low proportion of PDO crashes further emphasizes the vulnerability of pedestrians in traffic environments, reinforcing the critical importance of interventions aimed at reducing both fatal and injury-related incidents.

\subsection{Bus stops enviornment\label{sec4.2}}

Among the 596 bus stops analyzed, 280 (46.97\%) had no recorded pedestrian crashes during the study period. These non-crash locations were selected based on the inclusion criteria described in Subsection~\ref{sec4.1}. As illustrated in Fig.~\ref{Pic:2}(c), the distribution of crash frequency is heavily right-skewed, with a skewness value of 2.87, confirming the dominance of non-crash stops in the dataset..

This study distinguishes itself by integrating a diverse set of explanatory variables that capture spatial, geometric, and contextual dimensions often omitted in traditional modeling approaches. Table~\ref{table:1} presents descriptive statistics for the response variables and associated predictors. Data were compiled from multiple sources, including CRIS, RHiNo, Trinity Metro, Google Earth®, and Google Street View®, to ensure a comprehensive understanding of bus stop environments. The explanatory variables are organized into four overarching categories: exposure-related features, bus stop design characteristics, roadway attributes, and contextual elements. Exposure-related features reflect traffic and pedestrian activity near bus stops, incorporating metrics such as Average Annual Daily Traffic (AADT), boarding, and alighting counts. Bus stop design characteristics describe the physical layout and configuration of the stops, including design type and proximity to intersections. Roadway attributes capture infrastructure elements such as speed limits, lane counts, median type, and intersection distance. Contextual elements, such as land use type, lighting conditions, and nearby schools or parks, provide additional insight into the operating environment and pedestrian safety implications.

To evaluate the influence of crash severity, the crash data were stratified into three groups. The first group includes all pedestrian-involved crashes (KABCO). The second group (KABC) narrows the focus to fatal and injury crashes, excluding property damage only. The third group (KAB) concentrates on the most severe incidents, including fatalities and incapacitating or non-incapacitating injuries. This stratification enables a more nuanced investigation of the factors contributing to crash severity.

As shown in Table~\ref{table:1}, the mean number of crashes per bus stop is 0.752, with a maximum of 8 crashes at a single location. The mean KABC crash frequency is 0.526 (maximum = 6), while the KAB category shows a mean of 0.314 (maximum = 5). These values demonstrate considerable variation in crash occurrences across the study area, underscoring the importance of site-specific analysis.

In addition to crash data, several continuous variables were examined. AADT values ranged from 166 to 42056 vehicles per day, with a mean of 13558.2, reflecting the variation in traffic exposure near bus stops. Although direct pedestrian counts were unavailable, boarding and alighting volumes served as reasonable proxies for pedestrian activity. The distance to the nearest intersection ranged from 0.4 to 2106 feet, enabling analysis of spatial relationships between crashes and network geometry. Other continuous variables include median width, posted speed limit, number of lanes, number of nearby schools and parks, and the count of adjacent transit stops within a 0.25-mile radius. Several categorical variables were also included to account for environmental and design-related influences. These include intersection type (signalized vs. non-signalized), presence of crosswalks and sidewalks, roadway curvature, lighting conditions, area type (commercial, residential, or mixed), bus stop design (curbside vs. other), proximity of the stop to intersections (near, far, or midblock), and shelter availability. These features may collectively influence pedestrian behavior, visibility, and driver response, thus affecting crash risk. Table~\ref{table:1} summarizes these variables, illustrating the diverse conditions under which pedestrian crashes occur and providing a foundation for subsequent modeling and risk analysis.

\begin{table}[ht]
   \caption{Descriptive statistics of explanatory variables}
   \label{table:1}
   \centering
   \small
   \renewcommand{\arraystretch}{1.5}
   \resizebox{\textwidth}{!}
   {
    \begin{tabular}{l lcccc}
    \hline
    Variable & Definition &
    \multicolumn{4}{c}{Descriptive statistics}\\
    \multicolumn{1}{c}{} & &
    \multicolumn{1}{c}{Min} & Max&  Mean&Std Dev\\
    \hline
    Total crashes & KABCO frequency & 0 & 8 & 0.752 & 1.45 \\
    KABC crashes & KABC frequency  & 0 & 6 & 0.526 & 1.11 \\
    KAB crashes & KAB frequency & 0 & 5 & 0.314 & 0.65 \\
    \hline
    Continuous variables&&&&&\\
    \hline
    AADT & Traffic Volume on the street near bus stop & 166	& 42056 & 13540.6 &	8827.85 \\
    Avg on&	Average boarding&	0&	769.0&	53.20&	66.78\\
    Avg off&	Average alighting&	0&	831.0&	56.63&	60.64\\
    Dis to int&	Distance between bus stop and nearest intersection corner&	0.4&	2106&	198.94&	241.04\\
    Med w&	Median width&	0.00&	131.9&	7.10&	12.38\\
    Speed limit&	Posted speed limit&	20&	65&	36.63&	7.78\\
    Lane count&	Total number of lanes&	1&	8&	4.34&	1.46\\
    School&	Number of schools within a half-mile walk of the bus stop&	0&	6&	0.82&	1.12\\
    Park count&	Number of parks within a half-mile walk of the bus stop&	0&	6&	0.64&	0.84\\
    Stop count&	Count of transit stops (bus or rail) within a quarter mile of the bus stop&	0&	11&	3.45&	2.19\\   
    \hline
    Categorical variables&&Category&No of Stops&Crash Frequency\\
    \hline
    Int t&	Nearest intersection type&	Signalized (1)&	382&	237\\
		&&Non-signalized (0)	&214	&206\\
    Marked Xwalk&	Presence of marked crosswalk within 250ft of bus stop&	No (1)&	234	&176\\
		&&Yes (0)	&362	&267\\
    Med t&	Median types&	Undivided (1)&	311	&267\\
		&&Divided (0)&	297&	176\\
    Lighting&	Lighting presence&	No (1)&	339&	273\\
		&&Yes (0)&	257&	170\\
    Area&	Area type&	Com (0)&	307&	216\\
		&&Res (1)&	118&	116\\
		&&Mix (2)&	171&	111\\
    Sidewalks	&Presence of sidewalks	&Yes (0)	&550	&398\\
		&&No (1)	&46	&45\\
    Curve&	Presence of curve near bus stop&	Yes (1) &	38&	18\\
		&&No (0)&	558&	425\\
    Design&	Bus stop design&	Curbside (1)&	507&	416\\
		&&Other (0)&	89&	27\\
    Proximity&	The proximity of intersection from bus stop location&	Far (1)&	301&	207\\
		&&Near (0)	&226	&236\\
        &&Midblock (2)	&69&0\\
    Cover	&Presence of shelter&	Covered (0)	&302	&236\\
		&&Uncovered (1)	&294	&207\\
    \hline
    \end{tabular}%
   }
   \normalsize
\end{table}

\section{Results and Discussion} \label{sec5}
This section is divided into four subsections. First, it presents the model performance evaluation based on the KABCO crash severity scale. Second, it examines the key factors associated with pedestrian crashes near bus stops. Third, it explores the specific effects of these factors on severe crashes (KAB). Finally, the fourth subsection outlines the systematic identification and prioritization of high-risk bus stops using the proposed framework.

\subsection{Model performance evaluation \label{sec5.1}}

Although predictive accuracy was not the primary objective of this study, a preliminary assessment was conducted to evaluate the comparative performance of the models. Table \ref{table:2} summarizes these evaluation results, highlighting the RPNB-L model as superior, evidenced by its lowest DIC value. A difference of 10 or more in DIC values is generally regarded as significant (\citep{lord2021highway}); in this study, the RPNB-L model demonstrated a DIC value approximately 3.27\% lower than the NB-L model and 3.88\% lower than the RPNB-GE model, signifying a meaningful improvement in model performance.

Additionally, the RPNB-L model exhibited the lowest posterior mean deviance (Dbar), reflecting a better overall fit to the observed crash data compared to the competing models. Despite its higher complexity, indicated by the adequate number of parameters (PD), the RPNB-L model achieved a better balance between complexity and fit, resulting in superior predictive performance without substantial penalties. This balance underscores the model’s effectiveness in capturing unobserved heterogeneity and site-specific variability.

Regarding prediction accuracy, the RPNB-GE model displayed the highest MAE, signifying relatively less accurate predictions. Conversely, the RPNB-L model produced the lowest MAE for both training and testing datasets, as well as the lowest RMSE. Specifically, the RPNB-L model’s MAE and RMSE were approximately 3.98\% and 7.93\% lower, respectively, than the NB-L model with fixed coefficients. This superior predictive performance of the RPNB-L model can be attributed to its flexible structure, which accommodates mixed distributions and random parameters, thereby effectively capturing variations across bus stop locations. Consequently, the RPNB-L model emerges as the most robust and reliable analytical tool, offering enhanced accuracy and adaptability in modeling pedestrian crashes near bus stops.

\begin{table}[ht]
\caption{Posterior model estimated parameters for pedestrian-vehicle (KABCO) crashes near bus stops}
\label{table:2}
\centering
   \normalsize 
   \renewcommand{\arraystretch}{1.3} 
   \setlength{\tabcolsep}{5pt} 
   \resizebox{\textwidth}{!}{ 
    \begin{tabular}{p{5cm}ccccccccc}
    \hline
     & \multicolumn{3}{c}{\textbf{NB-L}}&\multicolumn{3}{c}{\textbf{RPNB-L}}&\multicolumn{3}{c}{\textbf{RPNB-GE}}\\
     \cline{2-4} \cline{5-7} \cline{8-10}
    {\textbf{Variables}}&Mean (Std. Dev.)&\multicolumn{2}{c}{95\% CI}&Mean (Std. Dev.)&\multicolumn{2}{c}{95\% CI} &Mean (Std. Dev.)&\multicolumn{2}{c}{95\% CI}\\
    \cline{3-4} \cline{6-7} \cline{9-10}
     & &LL&UL& &LL&UL&&LL&UL \\
    \hline 
    {\textbf{Mean of Parameters}} &&&&&&&&&\\  
    Constant& -0.776(0.076)&-0.879&	-0.578&	-0.488(0.031)&	-0.668&	-0.278 & -0.812 (0.08)&-0.898	&-0.545	\\
    {\textbf{Exposure variable}} &&&&&&&&&\\
    Ln (AADT)& 0.349(0.080)&	0.095&	0.411&	0.345(0.083)&	0.091&	0.424 &0.351(0.081) &0.096	&	0.040\\
    Avg on&	0.007(0.003)&	0.002&	0.013&	0.010(0.004)&	0.001&	0.014& 0.012(0.008)& 0.007	&	0.019\\
    {\textbf{Roadway environment}} &&&&&&&&&\\    

    Speed limit (1 if speed limit greater or equal to 35 mph, 0 otherwise)&	0.457(0.163)&	0.139&	0.778&	0.448(0.162)&	0.132&	0.760 &0.461(0.16)&0.141&0.791\\
    Med t (1 if nearest median is Undivided, 0 otherwise)&	0.321(0.149)&	0.027&	0.621&	0.326(0.148)&	0.0425&	0.624&0.323(0.150)&0.028&0.631\\
    Int t (1 if nearest intersection is signalized, 0 otherwise)&	-0.330(0.098)&	-0.622&	-0.004&	-0.329(0.108)&	-0.675&	-0.017&-0.334(0.010)&-0.624&-0.010\\
    {\textbf{Bus stop design features}} &&&&&&&&&\\
    Sidewalk (1 if sidewalk is present, 0 otherwise)&  	&	&	&	-0.379(0.235)&	-0.843&	-0.011&-0.446(0.312)&-0.976&-0.234\\  
    Marked Xwalk (1 if no presence of marked xwalk, 0 otherwise)& &	&	&	0.098 (0.075)&	0.002&	0.446&&&\\
    Lighting (1 if no presence of lighting, 0 otherwise)& 	&	&	&	0.100(0.014)&	0.060&	0.184&&&\\
    Area Mix (1 if bus stop located in mixed environment area, 0 otherwise)&  -0.355(0.161)&	-0.672&	-0.037&	-0.367(0.159)&	-0.642&	-0.029&-0.378(0.166)&-0.688&-0.039\\
    School count (1 if the Number of schools within a half-mile walk of the bus stop is greater than 1, 0 otherwise)& 0.562(0.181)&	0.272&	0.867&	0.556(0.179)&	0.200&	0.908 &&&\\
    Proximity (1 if bus stop located far side of intersection, 0 otherwise)&		-0.253(0.142)&	-0.483&	-0.023&	-0.251(0.141)&	-0.528&	-0.027&-0.252(0.141)&-0.471&-0.024\\

    {\textbf{Standard Deviation of Random Parameters}} &&&&&&&&&\\
    Ln (AADT) & 	&	&	&	0.042(0.005)&	0.044&	0.072&0.049(0.006)&0.046&0.089\\
    Avg On & 	&	&	&	0.008(0.001)&	0.003&	0.011&0.009(0.001)&0.003&0.012\\   

    Dispersion parameter ($\alpha$)&	0.365(0.09)&	0.124&	0.544&	0.137(0.083)&	0.067&	0.245&0.378(0.010)&0.112&0.576\\
    Lindley parameter ($\theta$)& 1.414(0.166)&	1.233&	1.766&	1.378(0.122)&	0.899&	1.589&&&\\

    GE parameter ($a$)&	&	&	&	&	&	&2.022(0.778)&1.089&2.532\\
    GE parameter ($b$)&	&	&	&	&	&	&1.442(0.378)&1.102&1.988\\
    
    \hline
    {\textbf{Performance measure}} &&&&&&\\
    \hline
    DIC&&	1233.56&&&\textbf{1193.23}&&&1241.4	\\
    Dbar&&	1114.06&&& \textbf{992.13}	&&&	1096.6\\
    Pd	&&119.5&&&\textbf{201.1}	&&&	144.8\\
    MAE (Train)	&&0.714&&&	\textbf{0.686}&&& 0.724	\\
    MAE (Test)	&&0.805&&&	\textbf{0.773}&&& 0.812	\\
    RMSE (Train)&&	0.956	&&&\textbf{0.881}&&&	0.968\\
    RMSE (Test)&&	1.072&&&	\textbf{0.987}	&&&1.104\\
    \hline
    \end{tabular}%
   }
   \normalsize
\end{table}

\begin{figure}[h!]
    \centering
    \setlength{\abovecaptionskip}{0pt}
    \subcaptionbox{AADT}
    {\includegraphics[width=0.6\textwidth]{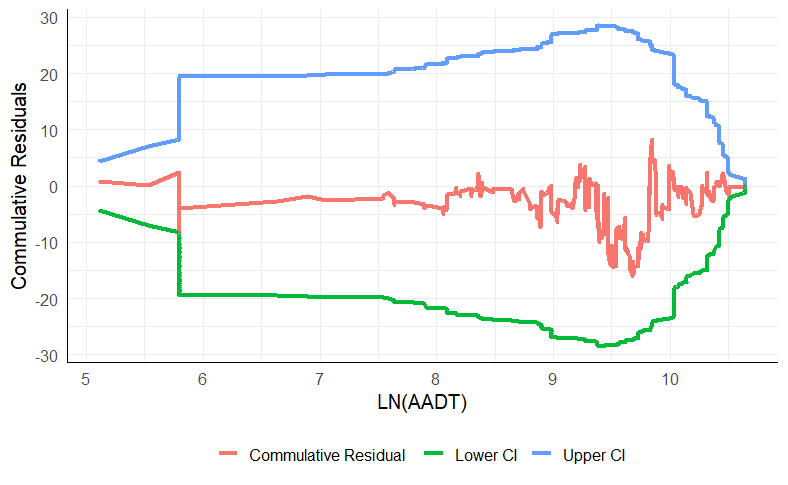}}
    \subcaptionbox{Avg On}{\includegraphics[width=0.6\textwidth]{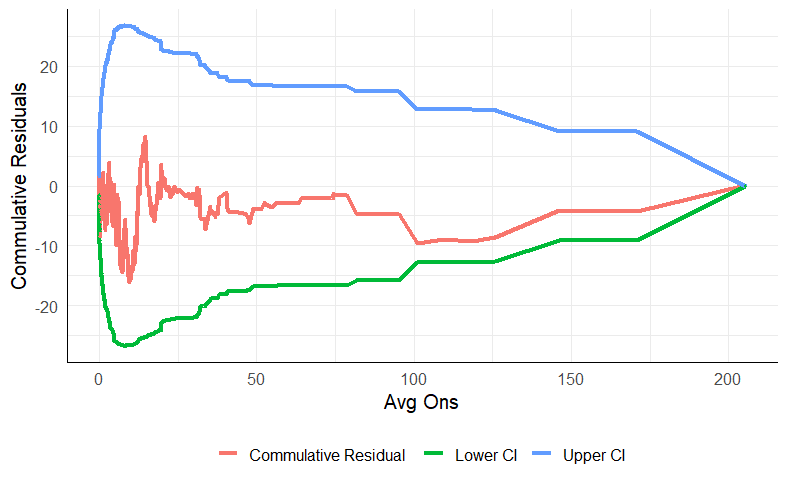}}
    \caption{CURE plots for RPNB-L model}
    \label{fig:7}
\end{figure}

Given the superior performance of the RPNB-L model across key metrics, an additional validation using the Cumulative Residual (CURE) plot was conducted to assess potential biases in model predictions. The CURE plot provides valuable insights into prediction accuracy by visually examining the accumulation of residuals across critical explanatory variables (Fig. \ref{fig:7}). Specifically, the cumulative residual behavior was evaluated for primary exposure variables, including Ln(AADT) and AvgOn. A desirable characteristic of the CURE plot is that cumulative residuals should consistently oscillate near the zero line, indicating unbiased predictions. However, in this analysis, the CURE plots exhibit prolonged smooth runs instead of frequent oscillations around the zero reference line. Rather than reflecting model overfitting, this pattern arises from the skewed distributions and uneven densities within the ranges of the exposure variables. Consequently, residuals accumulate smoothly over certain ranges of data points rather than fluctuating sharply. To further validate the assessment, 95\%  were computed by estimating residual variances over ascending values of explanatory variables and subsequently accumulating these variances. The cumulative residual lines for the RPNB-L model consistently remained close to the zero line, falling well within these confidence intervals. Overall, the CURE plot evaluation supports the robustness and predictive accuracy of the RPNB-L model, reinforcing its effectiveness in reliably modeling pedestrian crash risks near bus stops.

\subsection{Key contributing risk variables \label{sec5.3}}

Given the superior predictive performance demonstrated in Section \ref{sec5.1}, the RPNB-L model was selected for a more detailed interpretation and discussion. As previously described, the RPNB-L model incorporates random parameters within its hierarchical structure (see Eqn.\ref{Equation 13}), allowing the effects of explanatory variables to vary across bus stop locations. However, during the analysis, it became apparent that certain parameters exhibited minimal variability, contributing only negligible improvements to GOF and parameter mean estimates outside of 95\% CI. To balance model complexity with interpretability, random parameters were therefore assigned only to the primary exposure variables (e.g., Ln(AADT) and AvgOn) while the remaining parameters were treated as fixed (\citep{hasan2024short}). This approach simplified the model without sacrificing significant predictive power.

As summarized in Table \ref{table:2}, all parameter estimates remained statistically significant at the 95\% confidence interval, including the standard deviations for the randomly specified exposure parameters. To provide practical insights into the effects of these variables, average marginal effects were calculated, quantifying how variations in each explanatory variable influence crash frequency. The procedure for calculating marginal effects is detailed in Section \ref{sec3.6}, and the corresponding results are presented in Table \ref{table:3}.

\begin{table}[ht]
\caption{RPNB-L based average marginal effects of crash contributing parameters}
\label{table:3}
\centering
    \setlength{\tabcolsep}{10pt}
    \begin{tabular}{p{10cm}lc}
    \hline
    \textbf{Risk variables} & \textbf{Effects (Mean)}\\
    \hline
    Ln (AADT)&        	0.336\\
    Avg on&	0.018\\
    Speed limits (1 if speed limits greater or equal to 35 mph, 0 otherwise)&	0.259\\
    Med t (1 if nearest median is
    Undivided, 0 otherwise)&	0.245\\
    Int t (1 if nearest intersection is
    signalized, 0 otherwise)&	-0.247\\
    Sidewalk (1 if sidewalk is present,
    0 otherwise)& -0.284\\
    Marked Xwalk (1 if no presence of
    marked xwalk, 0 otherwise)& 0.074\\
    Lighting (1 if no presence of lighting , 0 otherwise)& 	0.075\\
        Area Mix (1 if bus stop is located in
    mixed environment area, 0 otherwise)&  -0.275\\
    Proximity (1 if bus stop is located far
    side of the intersection, 0 otherwise)&	-0.188\\
    School count (1 if the Number of schools within a half-mile walk of the bus stop is greater than 1, 0 otherwise) &	0.017\\
    \hline
    \end{tabular}%
\end{table}

\bigskip

\subsubsection{ Annual average daily traffic (AADT)}\label{sec5.3.1}

AADT near bus stops significantly influences pedestrian crash frequency. The RPNB-L model confirms that Ln(AADT) is statistically significant at the 95\% CI, indicating a positive relationship between traffic volume and crash frequency. Specifically, the estimated mean coefficient for Ln(AADT) is 0.345, suggesting that higher traffic volumes correspond to a greater likelihood of pedestrian crashes. This finding aligns with the expectation that increased vehicular flow elevates pedestrian exposure to collisions. Consistent with previous studies, these results highlight the risks associated with high traffic volumes near bus stops, which are often located in high-density passenger activity areas. Such areas experience significant traffic movement, increasing pedestrian exposure and crash risk (\citep{tiboni2013implementing}). The absence of dedicated bus lanes exacerbates this issue, as buses operating in mixed traffic negatively impact pedestrian safety (\citep{yendra2024comparison}). Additionally, studies have shown that higher AADT levels correlate with increased pedestrian fatalities in these environments (\citep{ulak2021stop,lakhotia2020pedestrian,hess2004pedestrian,amadori2012methodology}). The model's random parameterization for Ln(AADT), with a standard deviation of 0.042, highlights the variability in traffic volume impact across bus stop locations. This suggests that while traffic volume generally increases crash frequency, the magnitude of this effect varies between sites. Some bus stops may experience a steeper rise in crash risk with growing traffic volumes, while others may be less affected. This variability underscores the importance of considering site-specific factors in evaluating vehicular traffic's impact on pedestrian safety. Furthermore, the marginal effects of the RPNB-L model indicate that a one-unit increase in the logarithm of AADT corresponds to approximately a 0.336-unit increase in pedestrian-vehicle crash frequency near bus stops.

\subsubsection{Passengers density}\label{sec5.3.2}
In this study, pedestrian density near bus stops was represented by the mean number of passengers boarding and alighting at each stop, given the challenges of capturing real-time density. The model results indicate a positive association between higher pedestrian exposure, reflected in increased boarding activity, and crash frequency. Greater pedestrian volumes elevate the likelihood of vehicle-pedestrian interactions, increasing crash risk. The estimated mean coefficient for the "Avg on" variable is 0.010, with a standard deviation of 0.008, highlighting variability in effect across different bus stops due to factors such as traffic conditions and surrounding infrastructure. These findings are consistent with prior research (\citep{hess2004pedestrian, geedipally2021effects, craig2019pedestrian}), which links high pedestrian volumes near bus stops to elevated crash risks, particularly in dense areas like schools and commercial zones where large groups of pedestrians gather. The marginal effect indicates that an increase in average boarding activity raises passenger-vehicle crash frequencies by approximately 0.018 near bus stops. Both the model and previous studies emphasize that higher pedestrian density increases exposure to traffic, thereby raising crash risks.

\subsubsection{Speed limits} \label{sec5.3.3}
The results reveal a paradox: lower speed limits (35 mph or less) near bus stops are associated with increased pedestrian crashes. One potential cause is that pedestrians may feel safer with lower speed limits posted, which leads them to cross the road more frequently and increases their exposure to collisions. In addition, some drivers may exceed speed limits, resulting in unpredictable driving behaviors that further increase crash risks. Previous research has primarily focused on the dangers of high traffic speeds near bus stops, consistently linking higher speeds to an increased likelihood of pedestrian deaths (\citep{rossetti2020field, pulugurtha2008hazardous, hess2004pedestrian, lakhotia2020pedestrian}). As a result, lower speed limits have often been recommended to improve pedestrian safety (\citep{ulak2021stop, tubis2021method, pessaro2017impact}). However, the findings of this study suggest that simply posting reduced speed limits may not adequately mitigate crash risks. To effectively improve safety, stricter speed enforcement and enhanced pedestrian infrastructure, such as crosswalks, pedestrian signals, and traffic calming measures, are essential. These measures can help ensure driver compliance with speed limits while providing pedestrians with safe crossing options and reducing vehicle-pedestrian conflicts near bus stops.

\subsubsection{Median types}\label{sec5.3.4}
The findings show that roadways near bus stops without medians are significantly more likely to experience pedestrian crashes than those with medians. Medians serve as pedestrian refuges, allowing individuals to crossroads in stages and reducing their exposure to traffic. In contrast, undivided roadways require pedestrians to cross the entire width of the road at once, increasing crash risks due to unpredictable crossing behavior and limited driver visibility. These results align with previous studies highlighting the safety benefits of medians in reducing pedestrian-vehicle conflicts by providing protected spaces for pedestrians to pause mid-crossing (\citep{yu2024impact, salum2024toolkit, jeng2003pedestrian}). Medians offer an essential safety buffer, particularly on wide or high-traffic roads, by creating more predictable crossing points and minimizing pedestrian exposure to moving vehicles. The findings reinforce the importance of incorporating medians as a critical element in road design near bus stops to enhance pedestrian safety. Strategically placed medians ensure safer pedestrian movement and reduce potential conflicts with vehicles, ultimately contributing to lower crash risks (\citep{pessaro2017impact}).

\subsubsection{Nearest intersection}\label{sec5.3.5}
Intersections are widely recognized as critical zones for pedestrian risk due to the convergence of multimodal traffic streams and complex crossing behaviors. Numerous studies (\citep{harwood2008pedestrian, kucskapan2022pedestrian, walgren1998using, geedipally2021effects, quistberg2015bus}) have reported elevated pedestrian crash frequencies at or near bus stops located within 1000 feet of signalized intersections. However, the scope of these investigations has primarily focused on signalized intersections, often overlooking unsignalized intersections, which may present heightened risk due to the lack of formal traffic control mechanisms. To address this limitation, the present study examined the relationship between pedestrian crash risk and intersection type, specifically comparing signalized and unsignalized intersections. The RPNB-L model results (Table~\ref{table:2}) indicate a negative and statistically significant association between bus stops located near signalized intersections and pedestrian crashes (–0.329). This suggests that relative to unsignalized locations, signalized intersections may offer a protective effect by reducing crash likelihood. These findings align with those of Samani and Amador-Jimenez (\citeyear{samani2023exploring}), who observed that even modest infrastructure elements such as pedestrian signals and stop signs at non-signalized intersections can enhance driver awareness and pedestrian visibility. Signalized intersections likely reduce pedestrian risk by facilitating more orderly vehicle movement, minimizing turning conflicts, and providing designated pedestrian crossing phases.

To further validate these findings and explore potential confounding effects, a stratified Mantel–Haenszel (MH) analysis (\citep{santner2012statistical}) was conducted. This method provides an adjusted estimate of the association between crash occurrence and intersection type by accounting for key contextual variables across multiple strata. These strata were defined based on traffic volume (AADT), bus stop proximity to intersections, presence of marked crosswalks, and land use type. The MH estimator was computed using stratum-specific $2 \times 2$ contingency tables, where each table cross-tabulated pedestrian crash occurrence (yes/no) with intersection type (signalized/unsignalized). The pooled Mantel–Haenszel odds ratio (MHOR) is given by:

\begin{equation}
\hat{OR}_{MH} = \frac{\sum_{s=1}^S \frac{a_s d_s}{n_s}}{\sum_{s=1}^S \frac{b_s c_s}{n_s}}
\end{equation}

\noindent
where $a_s$, $b_s$, $c_s$, and $d_s$ represent the cell counts of the $s$-th stratum, and $n_s$ is the total number of observations within that stratum.

Results from the MH analysis corroborated the model-based findings. Signalized intersections were consistently associated with significantly lower crash risk across all stratified analyses. For example, when stratified by AADT with different ranges, the MHOR was 0.284, corresponding to a relative risk reduction of approximately 41\% (RR~$\approx$~0.59). Similarly, stratification by bus stop proximity on the far side yielded an MHOR of 0.313 (RR~$\approx$~0.62); by marked crosswalk presence, 0.272 (RR~$\approx$~0.56). All odds ratios were statistically significant at the $p < 0.001$ level. These consistent results across multiple strata highlight the presence of traffic control mechanisms, such as signal phases, pedestrian indications, and regulated turning movements, may contribute to reducing pedestrian exposure to conflicts and enhancing overall safety. Moreover, these findings help reconcile discrepancies in previous literature by illustrating how aggregated or unstratified analyses may obscure the protective influence of signalization.

\subsubsection{Sidewalks}\label{sec5.3.6}
The model results from Tables \ref{table:3} and \ref{table:4} show that the absence of sidewalks near bus stops significantly increases pedestrian crash likelihood, as indicated by the negative coefficient (-0.379). This finding suggests that without sidewalks, pedestrians are more exposed to traffic, often forced to walk on the road, thereby increasing the risk of vehicle-pedestrian collisions. Sidewalks provide a designated pedestrian space, minimizing traffic interactions and reducing crash risks. These results align with previous research emphasizing the critical role of sidewalks in enhancing pedestrian safety near bus stops. Pessaro et al. (\citeyear{pessaro2017impact}) highlighted that proper pedestrian infrastructure, such as sidewalks, encourages safer walking behavior. Similarly, Sukor and Fisal (\citeyear{sukor2020safety}) found that safety is the strongest factor influencing walkability to bus stops, further reinforcing the importance of adequate sidewalks. Tiboni and Rossetti (\citeyear{tiboni2013implementing}) found that one-third of the bus stops they surveyed lacked sidewalks, which exposed pedestrians to greater traffic risks. Likewise, Rossetti and Tiboni (\citeyear{rossetti2020field}) emphasized that pedestrians who walk along roadways without sidewalks face an increased risk of crashes. Overall, the findings highlight the critical importance of sidewalks in reducing pedestrian-vehicle interactions and ensuring safer access to bus stops (\citep{yu2024impact,samani2023exploring,lakhotia2020pedestrian,mukherjee2023built}), making sidewalks a vital component of pedestrian infrastructure.

\subsubsection{Marked crosswalks}\label{sec5.3.7}
The model results indicate (Table \ref{table:4}) that the absence of marked crosswalks near bus stops is associated with a higher risk of pedestrian crashes, as shown by approximately a 0.074 unit increase in pedestrian-vehicle crash frequency near bus stops. This finding suggests that locations without marked crosswalks are more likely to experience pedestrian crashes. Marked crosswalks provide a designated space for pedestrians to cross safely, reducing exposure to traffic and improving the predictability of pedestrian movements for drivers. Without crosswalks, pedestrians are more likely to cross at unpredictable locations, increasing crash risk due to reduced driver awareness and slower response times. Crosswalks also help organize traffic flow by signaling drivers to anticipate pedestrian activity, leading to safer road interactions. These findings align with previous research emphasizing the importance of marked pedestrian crossings near bus stops. For instance, Tiboni and Rossetti (\citeyear{tiboni2013implementing}) found that over 30\% of inspected bus stops lacked marked crossings, contributing to unsafe pedestrian conditions. Similarly, Mukherjee et al. (\citeyear{mukherjee2023built}) demonstrated that the presence of zebra markings at pedestrian crossings can reduce the likelihood of fatal crashes by 23\%, particularly during nighttime, by enhancing driver awareness of potential pedestrian interactions. The literature also highlights the value of pedestrian traffic signals in conjunction with crosswalks for improving safety. Studies show that pedestrian signals help regulate both driver and pedestrian behavior, encouraging drivers to expect pedestrians in crosswalks and prompting pedestrians to use designated areas safely (\citep{jeng2003pedestrian, samani2023exploring}). Without such infrastructure, pedestrians are more likely to take risks, particularly on wider roadways or bus stops lacking proper crossing facilities (\citep{lakhotia2020pedestrian}).

\subsubsection{Lighting}\label{sec5.3.8}
The inadequate lighting near bus stops significantly increases the likelihood of pedestrian crashes, as reflected by the positive coefficient of 0.100 in the RPNB-L model (Table \ref{table:3}). Poor lighting or no light at bus stops poses a higher risk during nighttime hours since its reduced visibility makes it difficult for pedestrians and drivers to anticipate movements. These findings align with previous research highlighting the critical role of adequate lighting in enhancing safety at bus stops. Mukherjee et al. (\citeyear{mukherjee2023built}) identified a direct relationship between insufficient lighting and higher pedestrian fatalities, especially in sparse land use. Similarly, Tiboni and Rossetti (\citeyear{tiboni2013implementing}) found that 8\% of the surveyed bus stops lacked artificial lighting, contributing to unsafe conditions. Studies consistently recommend enhanced illumination at bus stops and along their approaches to reduce crash risks (\citep{pessaro2017impact, salum2024toolkit}). Adequate lighting ensures that drivers and pedestrians have clear visibility, enabling them to anticipate movements and react appropriately, lowering crash frequency.  Overall, the model results and supporting literature underscore that improving lighting around bus stops is a critical safety measure for mitigating pedestrian crash risks.

\subsubsection{Area use}\label{sec5.3.9}

Previous research has identified bus stops in high pedestrian activity areas, mainly commercial zones with heavy traffic, as high-risk locations for pedestrian crashes. These elevated risks are often attributed to poor integration with surrounding land use and proximity to parking areas, which create additional conflict points between pedestrians and vehicles (\citep{pessaro2017impact, hess2004pedestrian}). The findings of this study also align with previous studies and suggest that bus stops located in mixed-use areas, combined with residential, commercial, and recreational spaces, exhibit a lower likelihood of pedestrian crashes than those in strictly commercial or residential zones. This reduction in crash risk may be attributed to balanced traffic flows and the presence of improved infrastructure, such as sidewalks, crosswalks, proper lighting, and pedestrian walking warning signs. In addition, enhanced safety measures like speed bumps and pedestrian signals in mixed-use areas further contribute to safer environments. Drivers in these zones may also exercise greater caution, given the diverse types of road users, including residents and recreational pedestrians.

\subsubsection{Bus stops proximity }\label{sec5.3.10}
The proximity of bus stops to intersections plays a critical role in pedestrian crash risks. Bus stops are categorized into three types based on their location: far-side, near-side, and mid-block. Far-side stops are located after intersections, near-side stops are positioned before intersections, and mid-block stops are between intersections. The model results indicate that far-side bus stops have a significantly lower likelihood of pedestrian crashes than near-side and mid-block stops, aligning with previous research (\citep{fitzpatrick1997effects, yu2024impact}). Far-side stops are considered safer because they reduce pedestrian-vehicle conflicts by allowing pedestrians to cross with greater visibility and ensuring that they cross behind the bus or at the nearest marked crosswalk. In contrast, near-side stops are associated with higher crash risks, likely due to reduced driver visibility and the tendency for pedestrians to cross in front of the bus, where drivers may not anticipate their movements. Similarly, mid-block stops pose increased risks because they often lack proper pedestrian infrastructure, such as marked crosswalks, leading to unpredictable crossing behavior (\citep{truong2011using, pessaro2017impact, rossetti2020field}).

\subsubsection{School count}\label{sec5.3.11}

The study demonstrates that the presence of schools near bus stops significantly increases the likelihood of pedestrian crashes. The results indicate that higher pedestrian density around schools, especially during peak times, leads to more pedestrian-vehicle interactions, thereby elevating the risk of crashes. While many students may be dropped off directly at school by buses, the presence of schools increases pedestrian activity around nearby bus stops, which raises the potential for conflicts between pedestrians and vehicles. These findings align with previous research, such as Craig et al. (\citeyear{craig2019pedestrian}) and Geedipally (\citeyear{geedipally2021effects}) found that bus stops near schools are associated with a higher incidence of collisions. Similarly, Hess et al. (\citeyear{hess2004pedestrian}) noted that gathering students around bus stops often leads to unpredictable pedestrian behavior, further heightening the risk of crashes.

\subsection{KAB crashes \label{sec5.4}}

The estimation using the RPNB-L model for KAB pedestrian crashes provides several key insights when compared to the total pedestrian crash (KABCO) model, as presented in Table \ref{table:4}. First, most significant risk factors identified in the KABCO model remain consistent in the KAB model. However, an exception was observed with marked crosswalks and sidewalks, which did not achieve statistical significance at the 95\% confidence level. confidence level in the KAB model. This suggests that these features may have a greater influence on less severe crashes, primarily reflected in the KABCO and KABC models. 

Second, similar to the KABCO model, traffic volume and average passenger boarding were found to be randomly distributed across locations in the KAB model. The posterior mean value for traffic volume (Ln AADT) was 0.191, with a standard deviation of 0.313, indicating its moderate effect on KAB crashes. Additionally, the coefficient for average passenger boarding (Avg on) was 0.007, with lower variability (standard deviation: 0.043). This suggests that while passenger boarding contributes to crash frequency, its impact on severe crashes is less pronounced. The coefficient for average passenger boarding further indicates that the relationship between traffic volume, speed limits, and pedestrian crash severity differs between severe (KAB) and less severe crashes. Specifically, lower traffic volumes on higher-speed roads tend to result in more severe crashes. The finding that higher speed limits (less or 35 mph) significantly increase crash severity reinforces this pattern, highlighting the critical role of speed in exacerbating the severity of pedestrian crashes near bus stops. The posterior parameter estimates further reveal that bus stops near signalized intersections experience fewer KAB crashes than those near non-signalized intersections. This underscores the importance of signalized intersections in reducing pedestrian crash risks, particularly for more severe incidents. Additionally, the absence of lighting at bus stops and proximity to schools were found to significantly increase crash severity. In contrast, far-side bus stops were associated with fewer severe crashes. These findings highlight the need for targeted safety interventions to address high-risk conditions and enhance pedestrian safety near bus stops.

\begin{table}[ht]
\caption{Posterior model estimated parameters for pedestrian-vehicle (KAB) crashes near bus stops}
\label{table:4}
\centering
\normalsize 
\renewcommand{\arraystretch}{1.2} 
\setlength{\tabcolsep}{7pt} 
\begin{tabular}{p{7cm}ccc}
\hline
 & \multicolumn{3}{c}{\textbf{KAB Crashes}} \\
\cline{2-4}

{\textbf{Variables}}&Mean (Std. Dev.)&\multicolumn{2}{c}{95\% CI}\\
    \cline{3-4}
     & &LL&UL\\
\hline 
\textbf{Mean of Parameters} & & & \\  
Constant & -0.453 (0.027) & -0.643 & -0.227 \\
\textbf{Exposure variable} & & & \\
Ln (AADT) & 0.191 (0.063) & 0.025 & 0.441 \\
Avg on & 0.007 (0.001) & 0.001 & 0.009 \\
\textbf{Roadway environment} & & & \\    
Speed limit (1 if speed limit $\geq$ 35 mph, 0 otherwise) & 0.393 (0.108) & 0.087 & 0.661 \\
Median type (1 if nearest median is undivided, 0 otherwise) & 0.217 (0.089) & 0.035 & 0.446 \\
Intersection type (1 if nearest intersection is signalized, 0 otherwise) & -0.292 (0.098) & -0.372 & -0.045 \\
\textbf{Bus stop design feature} & & & \\
Lighting (1 if no presence of lighting, 0 otherwise) & 0.231 (0.089) & 0.054 & 0.412 \\
Area Mix (1 if bus stop located in mixed environment, 0 otherwise) & -0.415 (0.186) & -0.917 & -0.029 \\
School count (1 if number of schools within 0.5 miles $\geq$ 1, 0 otherwise) & 0.641 (0.187) & 0.210 & 1.108 \\
Proximity (1 if bus stop located far side of intersection, 0 otherwise) & -0.312 (0.167) & -0.656 & -0.0617 \\
Dispersion parameter ($\alpha$) & 0.214 (0.118) & 0.119 & 0.692 \\
Lindley parameter ($\theta$) & 1.419 (0.168) & 1.244 & 1.778 \\
\textbf{Standard Deviation of Random Parameters} & & & \\
Ln (AADT) & 0.313 (0.154) & 0.078 & 0.546 \\
Avg on & 0.043 (0.022) & 0.010 & 0.132 \\    
\hline
\textbf{Performance Measure} & & & \\
\hline
DIC & &  869.56& \\
Dbar & &  765.06 &\\
Pd &  & 131.5 & \\
MAE (Train) &  & 0.449 &\\
MAE (Test) & &  0.505 &\\
RMSE (Train) &  & 0.577 &\\
RMSE (Test) &  & 0.676 &\\
\hline
\end{tabular}
\normalsize
\end{table}

\subsection{Hazard bus stops/corridors identification \label{sec5.5}}

Identification of hazardous bus stops, defined as locations with a higher expected number of crashes than similar stops (\cite{elvik2007state}), is crucial for authorities (\citep{hauer1984problem, persaud1999empirical,miranda2007bayesian,montella2010comparative}). These high-risk locations serve as focal points for prioritizing safety investments and help delineate hazard corridors. Hotspot analysis, a widely used approach for identifying these sites, ranks locations from highest to lowest risk (\citep{hauer2002screening,hauer2004statistical}). However, this process presents challenges due to random variability in crash occurrences, annual fluctuations, and the regression-to-the-mean (RTM) bias (\citep{hauer1984problem}). The RTM bias reflects the statistical tendency for extreme crash frequencies at a site to regress toward the long-term average over time due to natural variation. Failure to account for RTM bias can lead to misclassifying sites as hazardous based on temporary crash anomalies or overlooking genuinely high-risk locations during periods of unusually low crash activity, leading to inefficient safety resource allocation. To mitigate these challenges, expected crash frequencies are commonly used, as they adjust for RTM bias and provide a more reliable measure of risk (\citep{hauer1997observational}). These frequencies can be estimated using the Full Bayes (FB) and Empirical Bayes (EB) methods (\citep{miranda2007bayesian,montella2010comparative,khodadadi2022derivation}). The EB method, although widely used, has limitations in addressing unobserved heterogeneity, which can introduce bias in crash frequency predictions. Recent advancements, such as the simulation-based empirical Bayes (Sim-EB) approach, integrate random parameter models to better capture unobserved heterogeneity in crash risk estimation (\citep{tahir2022simulation}). This refinement enhances the estimation of crash modification factors, improving the precision of safety evaluations, particularly in before-and-after treatment studies.

However, the FB approach remains superior to both EB and Sim-EB methods, as it fully accounts for parameter uncertainty within a hierarchical Bayesian framework rather than relying solely on prior information from reference sites. By generating posterior samples without requiring closed-form solutions, FB provides greater accuracy in parameter estimation and predictions, particularly in complex models. Given these advantages, this study employs the FB method to ensure a comprehensive and precise identification of hazardous bus stops, establishing a robust framework for safety improvements. Expected crash frequencies were estimated, and hazardous bus stops were identified using the Potential for Safety Improvement (PSI) approach. PSI quantifies the difference between expected and predicted crashes, providing a structured measure of safety risk (\citep{hauer2002screening}). The expected crash frequency at each bus stop was calculated following Eqn \ref{eq22}, and for a more detailed understanding, readers are directed to \citep{khodadadi2022derivation}. Notably, the PSI method has yet to be applied using the RPNB-L model, presenting an opportunity for future research to explore this innovative approach.

\begin{equation}
p(y_{\text{exp}} \mid y_{\text{obs}}) = \int_{\eta, \phi} NB\left(e^{-\eta}, \phi\right) 
\left( \int_b p(\eta \mid b, y_{\text{obs}}) \, \pi_b \, db \right) \, \pi_\phi \, d\eta \, d\phi
\label{eq22}
\end{equation}

\begin{figure}[h!]
    \centering
    \setlength{\abovecaptionskip}{0pt}
    \subcaptionbox{Bus stop categories}
    {\includegraphics[width=0.6\textwidth]{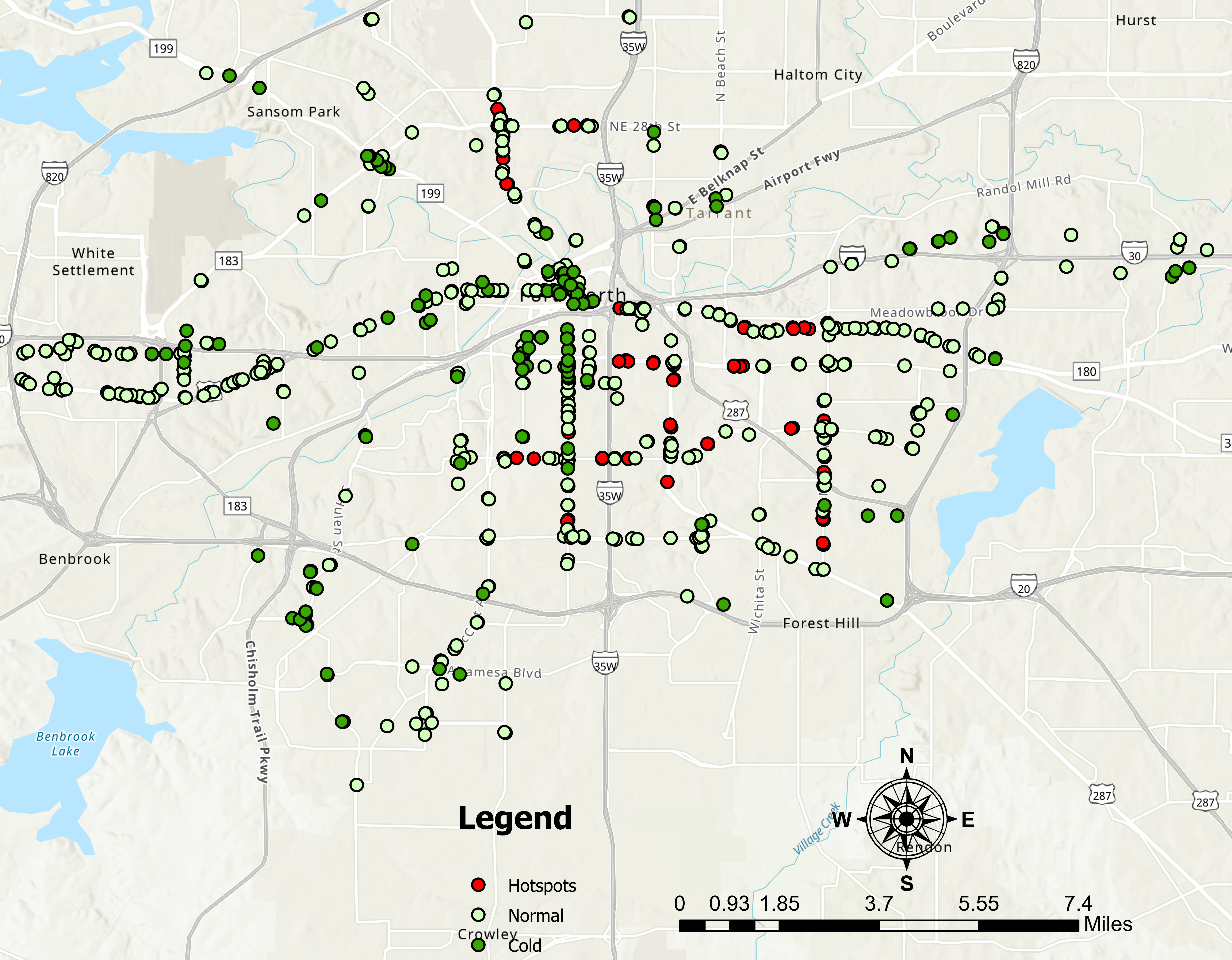}}
    \subcaptionbox{Hazard corridors}{\includegraphics[width=0.6\textwidth]{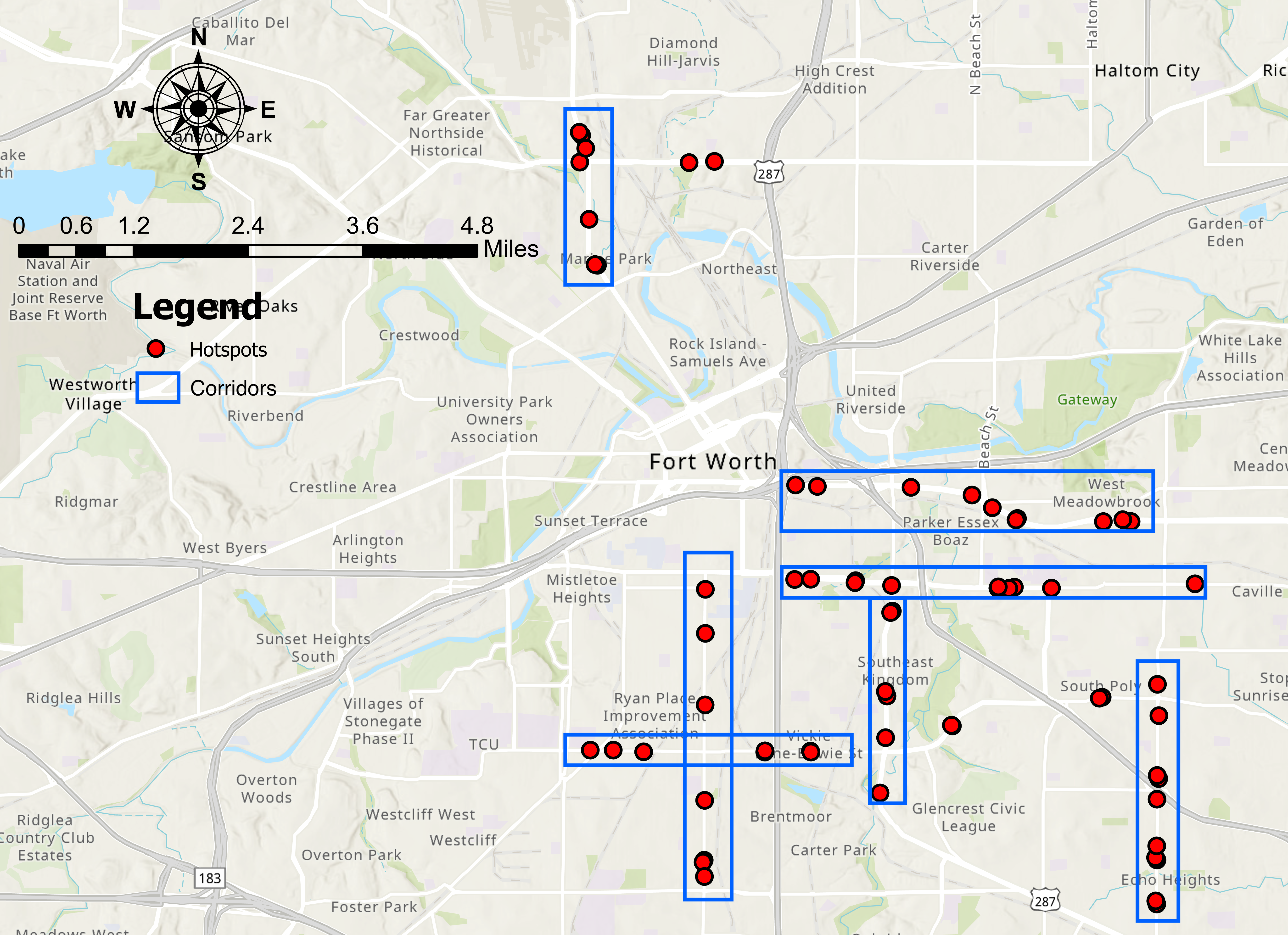}}
    \caption{Hazard bus stops and corridor identification}
    \label{fig:8}
\end{figure}

Based on PSI scores, bus stops were classified into three categories: (1) hotspots, where PSI values are positive and within the top 10\%, indicating high-priority locations for safety improvements; (2) normal zones, where PSI values are positive but not in the top 10\%, signaling locations that require attention but are less critical; and (3) cold zones, where PSI values are negative, suggesting relatively safer stops (see Fig. \ref{fig:8}a). This classification not only identifies hazardous bus stops but also provides a structured understanding of risk levels along transit corridors (see Fig. \ref{fig:8}b). Once high-risk corridors are identified, transportation agencies can prioritize interventions, ensuring effective resource allocation, particularly under budget constraints. 

\section{Summary and Conclusions}\label{sec 6}

This study highlights the growing concern regarding pedestrian safety near bus stops, emphasizing the necessity of identifying critical factors contributing to vehicle-pedestrian crashes in these environments. Previous research has primarily focused on pedestrian crashes at intersections and midblock locations, often treating bus stops merely as contributing factors rather than distinct safety assessment sites. Consequently, there remains a significant knowledge gap regarding the impacts of traffic exposure, roadway environment, and bus stop design on crash frequency near bus stops. Additionally, conventional statistical methods frequently overlook unobserved site-specific variations, potentially leading to ineffective safety interventions. In addition, it provides valuable insights by identifying hazardous bus stop locations, hotspots, and high-risk corridors. Ultimately, these findings are designed to support transportation agencies in implementing data-driven safety interventions, thereby enhancing the reliability and safety of urban transportation systems.

To address these gaps, the study utilized robust statistical modeling approaches, specifically employing the Random Parameters Negative Binomial-Lindley (RPNB-L) model, to analyze pedestrian crash data from 596 bus stops in Fort Worth, Texas, covering the period from 2018 to 2022. This model effectively captured unobserved heterogeneity and addressed the challenges associated with zero-crash observations and skewed crash data distributions, outperforming competing models (NB-L and RPNB-GE). 

Key factors identified as significantly influencing pedestrian crash occurrences included traffic volume (AADT), pedestrian boarding activity, intersection type, proximity to intersections, median type, presence of schools, speed limits, land use characteristics, sidewalk availability, lighting conditions, and the presence of marked crosswalks. Notably, bus stops near signalized intersections experienced fewer crashes due to controlled pedestrian crossings, whereas non-signalized intersections and near-side bus stops presented elevated crash risks. The presence of medians substantially improved pedestrian safety, facilitating safer crossings. Unexpectedly, zones with lower speed limits (35 mph or below) experienced higher crash frequencies, potentially due to increased pedestrian activity and perceived safety that encourages risky crossing behaviors. Similarly, locations near schools demonstrated elevated crash risks, while mixed-use areas were associated with fewer pedestrian incidents. Bus stop design elements, such as sidewalks, clearly marked crosswalks, and adequate lighting, significantly enhanced pedestrian safety. Proper lighting emerged as particularly critical during low-visibility conditions, while the absence of crosswalks markedly increased crash risks. Sidewalk availability reduced pedestrian exposure to vehicular traffic, thereby mitigating crash occurrences. Using the full Bays (FB) based the Potential for Safety Improvement (PSI) metric, the study systematically identified and prioritized hazardous bus stops and corridors, particularly in areas with changing roadway conditions, speed limits, or land-use patterns. The findings provide practical insights to transportation agencies for implementing targeted interventions, including infrastructure improvements, enhanced signage, speed management strategies, and public safety campaigns. Particular attention should be directed toward non-signalized intersections, areas with heavy pedestrian traffic, and school zones.

Despite these valuable insights, this study has limitations. Its geographic focus on Fort Worth, Texas, may restrict the broader applicability of findings. Additionally, reliance on historical crash data introduces potential inaccuracies and underreporting biases (\citep{arun2021systematic}). The absence of real-time pedestrian exposure data limits the precision of model predictions. Although random paramter techniques address unobserved heterogeneity, bias from omitted variables or measurement errors cannot be eliminated entirely. Additionally, this study primarily examines crashes occurring at bus stops, which could overlook risks encountered along pedestrian routes. Furthermore, variables such as weather conditions, law enforcement presence, pedestrian attitudes, and specific needs of vulnerable groups (children, older adults, and individuals with disabilities) were not considered, potentially impacting the comprehensiveness of the conclusions.

Future research directions include expanding the geographic scope to enhance model transferability by adding more urban areas and developing adjustment factors. Further studies should explore multivariate modeling structures, like the Random Parameters Poisson Lognormal-Lindley model (\citep{tahir2024poisson}), to better capture correlations between dependent variables and unobserved heterogeneity across crash severity levels (e.g., fatal and severe injury). Integrating real-time pedestrian exposure data, such as GPS tracking or pedestrian surveys, could provide a more dynamic understanding of behavior and risk. Future research should also focus on the safety needs of vulnerable populations, assessing the risks faced by children, older adults, and individuals with disabilities who rely heavily on public transportation. Additionally, pedestrian safety in low-density areas, where data are sparse and street vendor encroachment is a common issue, warrants further investigation. Additionally, investigating safety concerns in low-density and resource-constrained areas using innovative approaches, such as autonomous vehicle sensor data for real-time conflict detection (\citep{li2024beyond, anis2024real}), could significantly advance proactive pedestrian safety management offers a promising alternative to traditional crash data, particularly in resource-constrained settings. 

Overall, this study emphasizes the critical need to address pedestrian safety at bus stops systematically. The developed RPNB-L model provides an effective analytical framework to identify risk factors and hazardous locations, enabling targeted and data-driven safety improvements to enhance urban pedestrian safety and support reliable public transportation systems..

\bigskip

\subparagraph{\textbf{Acknowledgement:}}
The authors express their sincere gratitude to TxDOT for sponsoring the project from which this paper is derived. The authors thank TxDOT and Trinity Metro for providing the crash and bus stop data required for this paper.

\bigskip
\subparagraph{\textbf{Disclaimer:}}
The results presented in this document do not necessarily reflect those from the TxDOT.

\bigskip
\subparagraph{\textbf{Data availability:}}
Data and code will be made available on reasonable request.

\printcredits

\bibliographystyle{cas-model2-names}

\bibliography{ref}

\begin{thebibliography}{89}
\expandafter\ifx\csname natexlab\endcsname\relax\def\natexlab#1{#1}\fi
\providecommand{\url}[1]{\texttt{#1}}
\providecommand{\href}[2]{#2}
\providecommand{\path}[1]{#1}
\providecommand{\DOIprefix}{doi:}
\providecommand{\ArXivprefix}{arXiv:}
\providecommand{\URLprefix}{URL: }
\providecommand{\Pubmedprefix}{pmid:}
\providecommand{\doi}[1]{\href{http://dx.doi.org/#1}{\path{#1}}}
\providecommand{\Pubmed}[1]{\href{pmid:#1}{\path{#1}}}
\providecommand{\bibinfo}[2]{#2}
\ifx\xfnm\relax \def\xfnm[#1]{\unskip,\space#1}\fi
\bibitem[{Afghari et~al.(2014)Afghari, Ismail, Saunier, Sharma and Miranda-Moreno}]{afghari2014pedestrian}
\bibinfo{author}{Afghari, A.P.}, \bibinfo{author}{Ismail, K.}, \bibinfo{author}{Saunier, N.}, \bibinfo{author}{Sharma, A.}, \bibinfo{author}{Miranda-Moreno, L.}, \bibinfo{year}{2014}.
\newblock \bibinfo{title}{Pedestrian-cyclist interactions at bus stops along segregated bike paths: a case study of montreal}, in: \bibinfo{booktitle}{Transportation Research Board (USA) Annual Meeting}, pp. \bibinfo{pages}{1--1}.
\bibitem[{Akintayo and Adibeli(2022)}]{akintayo2022safety}
\bibinfo{author}{Akintayo, F.O.}, \bibinfo{author}{Adibeli, S.A.}, \bibinfo{year}{2022}.
\newblock \bibinfo{title}{Safety performance of selected bus stops in ibadan metropolis, nigeria}.
\newblock \bibinfo{journal}{Journal of Public Transportation} \bibinfo{volume}{24}, \bibinfo{pages}{100003}.
\bibitem[{Amadori and Bonino(2012)}]{amadori2012methodology}
\bibinfo{author}{Amadori, M.}, \bibinfo{author}{Bonino, T.}, \bibinfo{year}{2012}.
\newblock \bibinfo{title}{A methodology to define the level of safety of public transport bus stops, based on the concept of risk}.
\newblock \bibinfo{journal}{Procedia-Social and Behavioral Sciences} \bibinfo{volume}{48}, \bibinfo{pages}{653--662}.
\bibitem[{Anastasopoulos and Mannering(2009)}]{anastasopoulos2009note}
\bibinfo{author}{Anastasopoulos, P.C.}, \bibinfo{author}{Mannering, F.L.}, \bibinfo{year}{2009}.
\newblock \bibinfo{title}{A note on modeling vehicle accident frequencies with random-parameters count models}.
\newblock \bibinfo{journal}{Accident Analysis \& Prevention} \bibinfo{volume}{41}, \bibinfo{pages}{153--159}.
\bibitem[{Anis et~al.(2024)Anis, Li, Geedipally, Zhou and Lord}]{anis2024real}
\bibinfo{author}{Anis, M.}, \bibinfo{author}{Li, S.}, \bibinfo{author}{Geedipally, S.R.}, \bibinfo{author}{Zhou, Y.}, \bibinfo{author}{Lord, D.}, \bibinfo{year}{2024}.
\newblock \bibinfo{title}{Real-time risk estimation for active road safety: Leveraging waymo av sensor data with hierarchical bayesian extreme value models}.
\newblock \bibinfo{journal}{arXiv preprint arXiv:2407.16832} .
\bibitem[{Arun et~al.(2021)Arun, Haque, Bhaskar, Washington and Sayed}]{arun2021systematic}
\bibinfo{author}{Arun, A.}, \bibinfo{author}{Haque, M.M.}, \bibinfo{author}{Bhaskar, A.}, \bibinfo{author}{Washington, S.}, \bibinfo{author}{Sayed, T.}, \bibinfo{year}{2021}.
\newblock \bibinfo{title}{A systematic mapping review of surrogate safety assessment using traffic conflict techniques}.
\newblock \bibinfo{journal}{Accident Analysis \& Prevention} \bibinfo{volume}{153}, \bibinfo{pages}{106016}.
\bibitem[{Aryuyuen and Bodhisuwan(2013)}]{aryuyuen2013negative}
\bibinfo{author}{Aryuyuen, S.}, \bibinfo{author}{Bodhisuwan, W.}, \bibinfo{year}{2013}.
\newblock \bibinfo{title}{The negative binomial-generalized exponential (nb-ge) distribution}.
\newblock \bibinfo{journal}{Applied Mathematical Sciences} \bibinfo{volume}{7}, \bibinfo{pages}{1093--1105}.
\bibitem[{Barua et~al.(2016)Barua, El-Basyouny and Islam}]{barua2016multivariate}
\bibinfo{author}{Barua, S.}, \bibinfo{author}{El-Basyouny, K.}, \bibinfo{author}{Islam, M.T.}, \bibinfo{year}{2016}.
\newblock \bibinfo{title}{Multivariate random parameters collision count data models with spatial heterogeneity}.
\newblock \bibinfo{journal}{Analytic methods in accident research} \bibinfo{volume}{9}, \bibinfo{pages}{1--15}.
\bibitem[{Beck et~al.(2007)Beck, Dellinger and O'neil}]{beck2007motor}
\bibinfo{author}{Beck, L.F.}, \bibinfo{author}{Dellinger, A.M.}, \bibinfo{author}{O'neil, M.E.}, \bibinfo{year}{2007}.
\newblock \bibinfo{title}{Motor vehicle crash injury rates by mode of travel, united states: using exposure-based methods to quantify differences}.
\newblock \bibinfo{journal}{American Journal of Epidemiology} \bibinfo{volume}{166}, \bibinfo{pages}{212--218}.
\bibitem[{Campbell et~al.(2003)Campbell, Zegeer, Huang, Cynecki et~al.}]{campbell2003review}
\bibinfo{author}{Campbell, B.J.}, \bibinfo{author}{Zegeer, C.V.}, \bibinfo{author}{Huang, H.H.}, \bibinfo{author}{Cynecki, M.J.}, et~al., \bibinfo{year}{2003}.
\newblock \bibinfo{title}{A review of pedestrian safety research in the united states and abroad} .
\bibitem[{Chand and Chandra(2017)}]{chand2017improper}
\bibinfo{author}{Chand, S.}, \bibinfo{author}{Chandra, S.}, \bibinfo{year}{2017}.
\newblock \bibinfo{title}{Improper stopping of buses at curbside bus stops: Reasons and implications}.
\newblock \bibinfo{journal}{Transportation in developing economies} \bibinfo{volume}{3}, \bibinfo{pages}{1--9}.
\bibitem[{Clifton et~al.(2009)Clifton, Burnier and Akar}]{clifton2009severity}
\bibinfo{author}{Clifton, K.J.}, \bibinfo{author}{Burnier, C.V.}, \bibinfo{author}{Akar, G.}, \bibinfo{year}{2009}.
\newblock \bibinfo{title}{Severity of injury resulting from pedestrian--vehicle crashes: What can we learn from examining the built environment?}
\newblock \bibinfo{journal}{Transportation research part D: transport and environment} \bibinfo{volume}{14}, \bibinfo{pages}{425--436}.
\bibitem[{Craig et~al.(2019)Craig, Morris, Van~Houten and Mayou}]{craig2019pedestrian}
\bibinfo{author}{Craig, C.M.}, \bibinfo{author}{Morris, N.L.}, \bibinfo{author}{Van~Houten, R.}, \bibinfo{author}{Mayou, D.}, \bibinfo{year}{2019}.
\newblock \bibinfo{title}{Pedestrian safety and driver yielding near public transit stops}.
\newblock \bibinfo{journal}{Transportation research record} \bibinfo{volume}{2673}, \bibinfo{pages}{514--523}.
\bibitem[{Das et~al.(2019)Das, Dutta, Avelar, Dixon, Sun and Jalayer}]{das2019supervised}
\bibinfo{author}{Das, S.}, \bibinfo{author}{Dutta, A.}, \bibinfo{author}{Avelar, R.}, \bibinfo{author}{Dixon, K.}, \bibinfo{author}{Sun, X.}, \bibinfo{author}{Jalayer, M.}, \bibinfo{year}{2019}.
\newblock \bibinfo{title}{Supervised association rules mining on pedestrian crashes in urban areas: identifying patterns for appropriate countermeasures}.
\newblock \bibinfo{journal}{International Journal of Urban Sciences} \bibinfo{volume}{23}, \bibinfo{pages}{30--48}.
\bibitem[{El-Basyouny and Sayed(2009)}]{el2009urban}
\bibinfo{author}{El-Basyouny, K.}, \bibinfo{author}{Sayed, T.}, \bibinfo{year}{2009}.
\newblock \bibinfo{title}{Urban arterial accident prediction models with spatial effects}.
\newblock \bibinfo{journal}{Transportation research record} \bibinfo{volume}{2102}, \bibinfo{pages}{27--33}.
\bibitem[{Elvik(2007)}]{elvik2007state}
\bibinfo{author}{Elvik, R.}, \bibinfo{year}{2007}.
\newblock \bibinfo{title}{State-of-the-art approaches to road accident black spot management and safety analysis of road networks}.
\newblock \bibinfo{publisher}{Transport{\o}konomisk institutt Oslo, Norway}.
\bibitem[{Fitzpatrick and Nowlin(1997)}]{fitzpatrick1997effects}
\bibinfo{author}{Fitzpatrick, K.}, \bibinfo{author}{Nowlin, R.L.}, \bibinfo{year}{1997}.
\newblock \bibinfo{title}{Effects of bus stop design on suburban arterial operations}.
\newblock \bibinfo{journal}{Transportation Research Record} \bibinfo{volume}{1571}, \bibinfo{pages}{31--41}.
\bibitem[{Geedipally(2021)}]{geedipally2021effects}
\bibinfo{author}{Geedipally, S.}, \bibinfo{year}{2021}.
\newblock \bibinfo{title}{Effects of bus stops on pedestrian safety at signalized intersections}, in: \bibinfo{booktitle}{Conference of Transportation Research Group of India}, \bibinfo{organization}{Springer}. pp. \bibinfo{pages}{131--144}.
\bibitem[{Geedipally et~al.(2012)Geedipally, Lord and Dhavala}]{geedipally2012negative}
\bibinfo{author}{Geedipally, S.R.}, \bibinfo{author}{Lord, D.}, \bibinfo{author}{Dhavala, S.S.}, \bibinfo{year}{2012}.
\newblock \bibinfo{title}{The negative binomial-lindley generalized linear model: Characteristics and application using crash data}.
\newblock \bibinfo{journal}{Accident Analysis \& Prevention} \bibinfo{volume}{45}, \bibinfo{pages}{258--265}.
\bibitem[{Gelfand et~al.(1995)Gelfand, Sahu and Carlin}]{gelfand1995efficient}
\bibinfo{author}{Gelfand, A.E.}, \bibinfo{author}{Sahu, S.K.}, \bibinfo{author}{Carlin, B.P.}, \bibinfo{year}{1995}.
\newblock \bibinfo{title}{Efficient parametrisations for normal linear mixed models}.
\newblock \bibinfo{journal}{Biometrika} \bibinfo{volume}{82}, \bibinfo{pages}{479--488}.
\bibitem[{Gelman and Rubin(1992)}]{gelman1992inference}
\bibinfo{author}{Gelman, A.}, \bibinfo{author}{Rubin, D.B.}, \bibinfo{year}{1992}.
\newblock \bibinfo{title}{Inference from iterative simulation using multiple sequences}.
\newblock \bibinfo{journal}{Statistical science} \bibinfo{volume}{7}, \bibinfo{pages}{457--472}.
\bibitem[{Haleem et~al.(2015)Haleem, Alluri and Gan}]{haleem2015analyzing}
\bibinfo{author}{Haleem, K.}, \bibinfo{author}{Alluri, P.}, \bibinfo{author}{Gan, A.}, \bibinfo{year}{2015}.
\newblock \bibinfo{title}{Analyzing pedestrian crash injury severity at signalized and non-signalized locations}.
\newblock \bibinfo{journal}{Accident Analysis \& Prevention} \bibinfo{volume}{81}, \bibinfo{pages}{14--23}.
\bibitem[{Harwood et~al.(2008)Harwood, Bauer, Richard, Gilmore, Graham, Potts, Torbic and Hauer}]{harwood2008pedestrian}
\bibinfo{author}{Harwood, D.W.}, \bibinfo{author}{Bauer, K.M.}, \bibinfo{author}{Richard, K.R.}, \bibinfo{author}{Gilmore, D.K.}, \bibinfo{author}{Graham, J.L.}, \bibinfo{author}{Potts, I.B.}, \bibinfo{author}{Torbic, D.J.}, \bibinfo{author}{Hauer, E.}, \bibinfo{year}{2008}.
\newblock \bibinfo{title}{Pedestrian safety prediction methodology}.
\newblock \bibinfo{type}{Technical Report}.
\bibitem[{Hasan and Abdel-Aty(2024)}]{hasan2024short}
\bibinfo{author}{Hasan, T.}, \bibinfo{author}{Abdel-Aty, M.}, \bibinfo{year}{2024}.
\newblock \bibinfo{title}{Short-term safety performance functions by random parameters negative binomial-lindley model for part-time shoulder use}.
\newblock \bibinfo{journal}{Accident Analysis \& Prevention} \bibinfo{volume}{199}, \bibinfo{pages}{107498}.
\bibitem[{Hauer(1997)}]{hauer1997observational}
\bibinfo{author}{Hauer, E.}, \bibinfo{year}{1997}.
\newblock \bibinfo{title}{Observational before/after studies in road safety. Estimating the effect of highway and traffic engineering measures on road safety}.
\bibitem[{Hauer(2004)}]{hauer2004statistical}
\bibinfo{author}{Hauer, E.}, \bibinfo{year}{2004}.
\newblock \bibinfo{title}{Statistical road safety modeling}.
\newblock \bibinfo{journal}{Transportation Research Record} \bibinfo{volume}{1897}, \bibinfo{pages}{81--87}.
\bibitem[{Hauer et~al.(2002)Hauer, Kononov, Allery and Griffith}]{hauer2002screening}
\bibinfo{author}{Hauer, E.}, \bibinfo{author}{Kononov, J.}, \bibinfo{author}{Allery, B.}, \bibinfo{author}{Griffith, M.S.}, \bibinfo{year}{2002}.
\newblock \bibinfo{title}{Screening the road network for sites with promise}.
\newblock \bibinfo{journal}{Transportation Research Record} \bibinfo{volume}{1784}, \bibinfo{pages}{27--32}.
\bibitem[{Hauer and Persaud(1984)}]{hauer1984problem}
\bibinfo{author}{Hauer, E.}, \bibinfo{author}{Persaud, B.N.}, \bibinfo{year}{1984}.
\newblock \bibinfo{title}{Problem of identifying hazardous locations using accident data}.
\newblock \bibinfo{journal}{Transportation Research Record} \bibinfo{volume}{975}, \bibinfo{pages}{36--43}.
\bibitem[{Hess et~al.(2004)Hess, Moudon and Matlick}]{hess2004pedestrian}
\bibinfo{author}{Hess, P.M.}, \bibinfo{author}{Moudon, A.V.}, \bibinfo{author}{Matlick, J.M.}, \bibinfo{year}{2004}.
\newblock \bibinfo{title}{Pedestrian safety and transit corridors}.
\newblock \bibinfo{journal}{Journal of Public Transportation} \bibinfo{volume}{7}, \bibinfo{pages}{73--93}.
\bibitem[{on~Development of~the Highway Safety~Manual and on~the Highway Safety~Manual(2010)}]{national2010highway}
\bibinfo{author}{on~Development of~the Highway Safety~Manual, N.R.C.U.T.R.B.T.F.}, \bibinfo{author}{on~the Highway Safety~Manual, T.O.J.T.F.}, \bibinfo{year}{2010}.
\newblock \bibinfo{title}{Highway safety manual}. volume~\bibinfo{volume}{1}.
\newblock \bibinfo{publisher}{AASHTO}.
\bibitem[{Hu and Cicchino(2018)}]{hu2018examination}
\bibinfo{author}{Hu, W.}, \bibinfo{author}{Cicchino, J.B.}, \bibinfo{year}{2018}.
\newblock \bibinfo{title}{An examination of the increases in pedestrian motor-vehicle crash fatalities during 2009--2016}.
\newblock \bibinfo{journal}{Journal of safety research} \bibinfo{volume}{67}, \bibinfo{pages}{37--44}.
\bibitem[{Institute and Foundation(1996)}]{texas1996guidelines}
\bibinfo{author}{Institute, T.T.}, \bibinfo{author}{Foundation, T.A..M.R.}, \bibinfo{year}{1996}.
\newblock \bibinfo{title}{Guidelines for the location and design of bus stops}. volume~\bibinfo{volume}{19}.
\newblock \bibinfo{publisher}{National Academy Press}.
\bibitem[{Jeng et~al.(2003)Jeng, Fallat et~al.}]{jeng2003pedestrian}
\bibinfo{author}{Jeng, O.J.}, \bibinfo{author}{Fallat, G.}, et~al., \bibinfo{year}{2003}.
\newblock \bibinfo{title}{Pedestrian Safety and Mobility Aids for Crossings at Bus Stops}.
\newblock \bibinfo{type}{Technical Report}. New Jersey. Department of Transportation. Bureau of Research.
\bibitem[{Khan et~al.(2023)Khan, Afghari, Yasmin and Haque}]{khan2023effects}
\bibinfo{author}{Khan, S.A.}, \bibinfo{author}{Afghari, A.P.}, \bibinfo{author}{Yasmin, S.}, \bibinfo{author}{Haque, M.M.}, \bibinfo{year}{2023}.
\newblock \bibinfo{title}{Effects of design consistency on run-off-road crashes: an application of a random parameters negative binomial lindley model}.
\newblock \bibinfo{journal}{Accident Analysis \& Prevention} \bibinfo{volume}{186}, \bibinfo{pages}{107042}.
\bibitem[{Khodadadi et~al.(2023)Khodadadi, Shirazi, Geedipally and Lord}]{khodadadi2023evaluating}
\bibinfo{author}{Khodadadi, A.}, \bibinfo{author}{Shirazi, M.}, \bibinfo{author}{Geedipally, S.}, \bibinfo{author}{Lord, D.}, \bibinfo{year}{2023}.
\newblock \bibinfo{title}{Evaluating alternative variations of negative binomial--lindley distribution for modelling crash data}.
\newblock \bibinfo{journal}{Transportmetrica A: transport science} \bibinfo{volume}{19}, \bibinfo{pages}{2062480}.
\bibitem[{Khodadadi et~al.(2022)Khodadadi, Tsapakis, Shirazi, Das and Lord}]{khodadadi2022derivation}
\bibinfo{author}{Khodadadi, A.}, \bibinfo{author}{Tsapakis, I.}, \bibinfo{author}{Shirazi, M.}, \bibinfo{author}{Das, S.}, \bibinfo{author}{Lord, D.}, \bibinfo{year}{2022}.
\newblock \bibinfo{title}{Derivation of the empirical bayesian method for the negative binomial-lindley generalized linear model with application in traffic safety}.
\newblock \bibinfo{journal}{Accident Analysis \& Prevention} \bibinfo{volume}{170}, \bibinfo{pages}{106638}.
\bibitem[{Ku{\c{s}}kapan et~al.(2022)Ku{\c{s}}kapan, Sahraei, {\c{C}}odur and {\c{C}}odur}]{kucskapan2022pedestrian}
\bibinfo{author}{Ku{\c{s}}kapan, E.}, \bibinfo{author}{Sahraei, M.A.}, \bibinfo{author}{{\c{C}}odur, M.K.}, \bibinfo{author}{{\c{C}}odur, M.Y.}, \bibinfo{year}{2022}.
\newblock \bibinfo{title}{Pedestrian safety at signalized intersections: Spatial and machine learning approaches}.
\newblock \bibinfo{journal}{Journal of Transport \& Health} \bibinfo{volume}{24}, \bibinfo{pages}{101322}.
\bibitem[{Lakhotia et~al.(2020)Lakhotia, Lassarre, Rao and Tiwari}]{lakhotia2020pedestrian}
\bibinfo{author}{Lakhotia, S.}, \bibinfo{author}{Lassarre, S.}, \bibinfo{author}{Rao, K.R.}, \bibinfo{author}{Tiwari, G.}, \bibinfo{year}{2020}.
\newblock \bibinfo{title}{Pedestrian accessibility and safety around bus stops in delhi}.
\newblock \bibinfo{journal}{IATSS research} \bibinfo{volume}{44}, \bibinfo{pages}{55--66}.
\bibitem[{Lee et~al.(2015)Lee, Abdel-Aty, Choi and Huang}]{lee2015multi}
\bibinfo{author}{Lee, J.}, \bibinfo{author}{Abdel-Aty, M.}, \bibinfo{author}{Choi, K.}, \bibinfo{author}{Huang, H.}, \bibinfo{year}{2015}.
\newblock \bibinfo{title}{Multi-level hot zone identification for pedestrian safety}.
\newblock \bibinfo{journal}{Accident Analysis \& Prevention} \bibinfo{volume}{76}, \bibinfo{pages}{64--73}.
\bibitem[{Li et~al.(2017)Li, Ranjitkar, Zhao, Yi and Rashidi}]{li2017analyzing}
\bibinfo{author}{Li, D.}, \bibinfo{author}{Ranjitkar, P.}, \bibinfo{author}{Zhao, Y.}, \bibinfo{author}{Yi, H.}, \bibinfo{author}{Rashidi, S.}, \bibinfo{year}{2017}.
\newblock \bibinfo{title}{Analyzing pedestrian crash injury severity under different weather conditions}.
\newblock \bibinfo{journal}{Traffic injury prevention} \bibinfo{volume}{18}, \bibinfo{pages}{427--430}.
\bibitem[{Li et~al.(2024)Li, Anis, Lord, Zhang, Zhou and Ye}]{li2024beyond}
\bibinfo{author}{Li, S.}, \bibinfo{author}{Anis, M.}, \bibinfo{author}{Lord, D.}, \bibinfo{author}{Zhang, H.}, \bibinfo{author}{Zhou, Y.}, \bibinfo{author}{Ye, X.}, \bibinfo{year}{2024}.
\newblock \bibinfo{title}{Beyond 1d and oversimplified kinematics: A generic analytical framework for surrogate safety measures}.
\newblock \bibinfo{journal}{Accident Analysis \& Prevention} \bibinfo{volume}{204}, \bibinfo{pages}{107649}.
\bibitem[{Lord and Geedipally(2011)}]{lord2011negative}
\bibinfo{author}{Lord, D.}, \bibinfo{author}{Geedipally, S.R.}, \bibinfo{year}{2011}.
\newblock \bibinfo{title}{The negative binomial--lindley distribution as a tool for analyzing crash data characterized by a large amount of zeros}.
\newblock \bibinfo{journal}{Accident Analysis \& Prevention} \bibinfo{volume}{43}, \bibinfo{pages}{1738--1742}.
\bibitem[{Lord et~al.(2021)Lord, Qin and Geedipally}]{lord2021highway}
\bibinfo{author}{Lord, D.}, \bibinfo{author}{Qin, X.}, \bibinfo{author}{Geedipally, S.R.}, \bibinfo{year}{2021}.
\newblock \bibinfo{title}{Highway safety analytics and modeling}.
\newblock \bibinfo{publisher}{Elsevier}.
\bibitem[{Lord et~al.(2007)Lord, Washington and Ivan}]{lord2007further}
\bibinfo{author}{Lord, D.}, \bibinfo{author}{Washington, S.}, \bibinfo{author}{Ivan, J.N.}, \bibinfo{year}{2007}.
\newblock \bibinfo{title}{Further notes on the application of zero-inflated models in highway safety}.
\newblock \bibinfo{journal}{Accident Analysis \& Prevention} \bibinfo{volume}{39}, \bibinfo{pages}{53--57}.
\bibitem[{Lord et~al.(2005)Lord, Washington and Ivan}]{lord2005poisson}
\bibinfo{author}{Lord, D.}, \bibinfo{author}{Washington, S.P.}, \bibinfo{author}{Ivan, J.N.}, \bibinfo{year}{2005}.
\newblock \bibinfo{title}{Poisson, poisson-gamma and zero-inflated regression models of motor vehicle crashes: balancing statistical fit and theory}.
\newblock \bibinfo{journal}{Accident Analysis \& Prevention} \bibinfo{volume}{37}, \bibinfo{pages}{35--46}.
\bibitem[{Mannering et~al.(2016)Mannering, Shankar and Bhat}]{mannering2016unobserved}
\bibinfo{author}{Mannering, F.L.}, \bibinfo{author}{Shankar, V.}, \bibinfo{author}{Bhat, C.R.}, \bibinfo{year}{2016}.
\newblock \bibinfo{title}{Unobserved heterogeneity and the statistical analysis of highway accident data}.
\newblock \bibinfo{journal}{Analytic methods in accident research} \bibinfo{volume}{11}, \bibinfo{pages}{1--16}.
\bibitem[{Miranda-Moreno et~al.(2007)Miranda-Moreno, Labbe and Fu}]{miranda2007bayesian}
\bibinfo{author}{Miranda-Moreno, L.F.}, \bibinfo{author}{Labbe, A.}, \bibinfo{author}{Fu, L.}, \bibinfo{year}{2007}.
\newblock \bibinfo{title}{Bayesian multiple testing procedures for hotspot identification}.
\newblock \bibinfo{journal}{Accident Analysis \& Prevention} \bibinfo{volume}{39}, \bibinfo{pages}{1192--1201}.
\bibitem[{Mitra and Washington(2007)}]{mitra2007nature}
\bibinfo{author}{Mitra, S.}, \bibinfo{author}{Washington, S.}, \bibinfo{year}{2007}.
\newblock \bibinfo{title}{On the nature of over-dispersion in motor vehicle crash prediction models}.
\newblock \bibinfo{journal}{Accident Analysis \& Prevention} \bibinfo{volume}{39}, \bibinfo{pages}{459--468}.
\bibitem[{Montella(2010)}]{montella2010comparative}
\bibinfo{author}{Montella, A.}, \bibinfo{year}{2010}.
\newblock \bibinfo{title}{A comparative analysis of hotspot identification methods}.
\newblock \bibinfo{journal}{Accident Analysis \& Prevention} \bibinfo{volume}{42}, \bibinfo{pages}{571--581}.
\bibitem[{Mukherjee et~al.(2023)Mukherjee, Rao and Tiwari}]{mukherjee2023built}
\bibinfo{author}{Mukherjee, D.}, \bibinfo{author}{Rao, K.R.}, \bibinfo{author}{Tiwari, G.}, \bibinfo{year}{2023}.
\newblock \bibinfo{title}{Built-environment risk assessment for pedestrians near bus-stops: a case study in delhi}.
\newblock \bibinfo{journal}{International journal of injury control and safety promotion} \bibinfo{volume}{30}, \bibinfo{pages}{185--194}.
\bibitem[{Persaud et~al.(1999)Persaud, Lyon and Nguyen}]{persaud1999empirical}
\bibinfo{author}{Persaud, B.}, \bibinfo{author}{Lyon, C.}, \bibinfo{author}{Nguyen, T.}, \bibinfo{year}{1999}.
\newblock \bibinfo{title}{Empirical bayes procedure for ranking sites for safety investigation by potential for safety improvement}.
\newblock \bibinfo{journal}{Transportation research record} \bibinfo{volume}{1665}, \bibinfo{pages}{7--12}.
\bibitem[{Pessaro et~al.(2017)Pessaro, Catal{\'a}, Wang, Spicer et~al.}]{pessaro2017impact}
\bibinfo{author}{Pessaro, B.}, \bibinfo{author}{Catal{\'a}, M.}, \bibinfo{author}{Wang, Z.}, \bibinfo{author}{Spicer, M.}, et~al., \bibinfo{year}{2017}.
\newblock \bibinfo{title}{Impact of transit stop location on pedestrian safety} .
\bibitem[{Pulugurtha and Sambhara(2011)}]{pulugurtha2011pedestrian}
\bibinfo{author}{Pulugurtha, S.S.}, \bibinfo{author}{Sambhara, V.R.}, \bibinfo{year}{2011}.
\newblock \bibinfo{title}{Pedestrian crash estimation models for signalized intersections}.
\newblock \bibinfo{journal}{Accident Analysis \& Prevention} \bibinfo{volume}{43}, \bibinfo{pages}{439--446}.
\bibitem[{Pulugurtha and Vanapalli(2008)}]{pulugurtha2008hazardous}
\bibinfo{author}{Pulugurtha, S.S.}, \bibinfo{author}{Vanapalli, V.K.}, \bibinfo{year}{2008}.
\newblock \bibinfo{title}{Hazardous bus stops identification: An illustration using gis}.
\newblock \bibinfo{journal}{Journal of Public Transportation} \bibinfo{volume}{11}, \bibinfo{pages}{65--83}.
\bibitem[{Quistberg et~al.(2015a)Quistberg, Howard, Ebel, Moudon, Saelens, Hurvitz, Curtin and Rivara}]{quistberg2015multilevel}
\bibinfo{author}{Quistberg, D.A.}, \bibinfo{author}{Howard, E.J.}, \bibinfo{author}{Ebel, B.E.}, \bibinfo{author}{Moudon, A.V.}, \bibinfo{author}{Saelens, B.E.}, \bibinfo{author}{Hurvitz, P.M.}, \bibinfo{author}{Curtin, J.E.}, \bibinfo{author}{Rivara, F.P.}, \bibinfo{year}{2015}a.
\newblock \bibinfo{title}{Multilevel models for evaluating the risk of pedestrian--motor vehicle collisions at intersections and mid-blocks}.
\newblock \bibinfo{journal}{Accident Analysis \& Prevention} \bibinfo{volume}{84}, \bibinfo{pages}{99--111}.
\bibitem[{Quistberg et~al.(2015b)Quistberg, Koepsell, Johnston, Boyle, Miranda and Ebel}]{quistberg2015bus}
\bibinfo{author}{Quistberg, D.A.}, \bibinfo{author}{Koepsell, T.D.}, \bibinfo{author}{Johnston, B.D.}, \bibinfo{author}{Boyle, L.N.}, \bibinfo{author}{Miranda, J.J.}, \bibinfo{author}{Ebel, B.E.}, \bibinfo{year}{2015}b.
\newblock \bibinfo{title}{Bus stops and pedestrian--motor vehicle collisions in lima, peru: a matched case--control study}.
\newblock \bibinfo{journal}{Injury prevention} \bibinfo{volume}{21}, \bibinfo{pages}{e15--e22}.
\bibitem[{Rahman~Shaon and Qin(2016)}]{rahman2016use}
\bibinfo{author}{Rahman~Shaon, M.R.}, \bibinfo{author}{Qin, X.}, \bibinfo{year}{2016}.
\newblock \bibinfo{title}{Use of mixed distribution generalized linear models to quantify safety effects of rural roadway features}.
\newblock \bibinfo{journal}{Transportation Research Record} \bibinfo{volume}{2583}, \bibinfo{pages}{134--141}.
\bibitem[{Rossetti et~al.(2020)Rossetti, Tiboni et~al.}]{rossetti2020field}
\bibinfo{author}{Rossetti, S.}, \bibinfo{author}{Tiboni, M.}, et~al., \bibinfo{year}{2020}.
\newblock \bibinfo{title}{In field assessment of safety, security, comfort and accessibility of bus stops: A planning perspective}.
\newblock \bibinfo{journal}{EUROPEAN TRANSPORT/TRASPORTI EUROPEI} .
\bibitem[{Rusli et~al.(2018)Rusli, Haque, Afghari and King}]{rusli2018applying}
\bibinfo{author}{Rusli, R.}, \bibinfo{author}{Haque, M.M.}, \bibinfo{author}{Afghari, A.P.}, \bibinfo{author}{King, M.}, \bibinfo{year}{2018}.
\newblock \bibinfo{title}{Applying a random parameters negative binomial lindley model to examine multi-vehicle crashes along rural mountainous highways in malaysia}.
\newblock \bibinfo{journal}{Accident Analysis \& Prevention} \bibinfo{volume}{119}, \bibinfo{pages}{80--90}.
\bibitem[{Salum et~al.(2024)Salum, Kisabanzira, Alluri and Sando}]{salum2024toolkit}
\bibinfo{author}{Salum, J.H.}, \bibinfo{author}{Kisabanzira, N.}, \bibinfo{author}{Alluri, P.}, \bibinfo{author}{Sando, T.}, \bibinfo{year}{2024}.
\newblock \bibinfo{title}{Toolkit for the assessment of safety at bus stops}.
\newblock \bibinfo{journal}{Transportation research record} \bibinfo{volume}{2678}, \bibinfo{pages}{279--292}.
\bibitem[{Samani and Amador-Jimenez(2023)}]{samani2023exploring}
\bibinfo{author}{Samani, R.R.}, \bibinfo{author}{Amador-Jimenez, L.}, \bibinfo{year}{2023}.
\newblock \bibinfo{title}{Exploring road safety of pedestrians in proximity to public transit access points (bus stops and metro stations), a case study of montreal, canada}.
\newblock \bibinfo{journal}{Canadian Journal of Civil Engineering} \bibinfo{volume}{50}, \bibinfo{pages}{536--547}.
\bibitem[{Sanchez~Rodriguez and Ferenchak(2024)}]{sanchez2024longitudinal}
\bibinfo{author}{Sanchez~Rodriguez, O.}, \bibinfo{author}{Ferenchak, N.N.}, \bibinfo{year}{2024}.
\newblock \bibinfo{title}{Longitudinal spatial trends in us pedestrian fatalities, 1999--2020}.
\newblock \bibinfo{journal}{Transportation research record} \bibinfo{volume}{2678}, \bibinfo{pages}{904--916}.
\bibitem[{Santner and Duffy(2012)}]{santner2012statistical}
\bibinfo{author}{Santner, T.J.}, \bibinfo{author}{Duffy, D.E.}, \bibinfo{year}{2012}.
\newblock \bibinfo{title}{The statistical analysis of discrete data}.
\newblock \bibinfo{publisher}{Springer Science \& Business Media}.
\bibitem[{Shaon et~al.(2018)Shaon, Qin, Shirazi, Lord and Geedipally}]{shaon2018developing}
\bibinfo{author}{Shaon, M.R.R.}, \bibinfo{author}{Qin, X.}, \bibinfo{author}{Shirazi, M.}, \bibinfo{author}{Lord, D.}, \bibinfo{author}{Geedipally, S.R.}, \bibinfo{year}{2018}.
\newblock \bibinfo{title}{Developing a random parameters negative binomial-lindley model to analyze highly over-dispersed crash count data}.
\newblock \bibinfo{journal}{Analytic methods in accident research} \bibinfo{volume}{18}, \bibinfo{pages}{33--44}.
\bibitem[{Shinar(2012)}]{shinar2012safety}
\bibinfo{author}{Shinar, D.}, \bibinfo{year}{2012}.
\newblock \bibinfo{title}{Safety and mobility of vulnerable road users: pedestrians, bicyclists, and motorcyclists}.
\bibitem[{Stewart(2023)}]{stewart2023overview}
\bibinfo{author}{Stewart, T.}, \bibinfo{year}{2023}.
\newblock \bibinfo{title}{Overview of motor vehicle traffic crashes in 2021}.
\newblock \bibinfo{type}{Technical Report}.
\bibitem[{Sukor and Fisal(2020)}]{sukor2020safety}
\bibinfo{author}{Sukor, N.S.A.}, \bibinfo{author}{Fisal, S.F.M.}, \bibinfo{year}{2020}.
\newblock \bibinfo{title}{Safety, connectivity, and comfortability as improvement indicators of walkability to the bus stops in penang island}.
\newblock \bibinfo{journal}{Engineering, Technology \& Applied Science Research} \bibinfo{volume}{10}, \bibinfo{pages}{6450--6455}.
\bibitem[{Tahir et~al.(2022)Tahir, Haque, Yasmin and King}]{tahir2022simulation}
\bibinfo{author}{Tahir, H.B.}, \bibinfo{author}{Haque, M.M.}, \bibinfo{author}{Yasmin, S.}, \bibinfo{author}{King, M.}, \bibinfo{year}{2022}.
\newblock \bibinfo{title}{A simulation-based empirical bayes approach: Incorporating unobserved heterogeneity in the before-after evaluation of engineering treatments}.
\newblock \bibinfo{journal}{Accident Analysis \& Prevention} \bibinfo{volume}{165}, \bibinfo{pages}{106527}.
\bibitem[{Tahir et~al.(2024)Tahir, Yasmin and Haque}]{tahir2024poisson}
\bibinfo{author}{Tahir, H.B.}, \bibinfo{author}{Yasmin, S.}, \bibinfo{author}{Haque, M.M.}, \bibinfo{year}{2024}.
\newblock \bibinfo{title}{A poisson lognormal-lindley model for simultaneous estimation of multiple crash-types: Application of multivariate and pooled univariate models}.
\newblock \bibinfo{journal}{Analytic Methods in Accident Research} \bibinfo{volume}{41}, \bibinfo{pages}{100315}.
\bibitem[{Thakali et~al.(2015)Thakali, Kwon and Fu}]{thakali2015identification}
\bibinfo{author}{Thakali, L.}, \bibinfo{author}{Kwon, T.J.}, \bibinfo{author}{Fu, L.}, \bibinfo{year}{2015}.
\newblock \bibinfo{title}{Identification of crash hotspots using kernel density estimation and kriging methods: a comparison}.
\newblock \bibinfo{journal}{Journal of Modern Transportation} \bibinfo{volume}{23}, \bibinfo{pages}{93--106}.
\bibitem[{Tiboni and Rossetti(2013)}]{tiboni2013implementing}
\bibinfo{author}{Tiboni, M.}, \bibinfo{author}{Rossetti, S.}, \bibinfo{year}{2013}.
\newblock \bibinfo{title}{Implementing a road safety review approach for existing bus stops}.
\newblock \bibinfo{journal}{WIT Transactions on the Built Environment} \bibinfo{volume}{130}, \bibinfo{pages}{699--709}.
\bibitem[{Toran~Pour et~al.(2017)Toran~Pour, Moridpour, Tay and Rajabifard}]{toran2017modelling}
\bibinfo{author}{Toran~Pour, A.}, \bibinfo{author}{Moridpour, S.}, \bibinfo{author}{Tay, R.}, \bibinfo{author}{Rajabifard, A.}, \bibinfo{year}{2017}.
\newblock \bibinfo{title}{Modelling pedestrian crash severity at mid-blocks}.
\newblock \bibinfo{journal}{Transportmetrica A: Transport Science} \bibinfo{volume}{13}, \bibinfo{pages}{273--297}.
\bibitem[{Torbic et~al.(2010)Torbic, Harwood, Bokenkroger, Srinivasan, Carter, Zegeer and Lyon}]{torbic2010pedestrian}
\bibinfo{author}{Torbic, D.J.}, \bibinfo{author}{Harwood, D.W.}, \bibinfo{author}{Bokenkroger, C.D.}, \bibinfo{author}{Srinivasan, R.}, \bibinfo{author}{Carter, D.}, \bibinfo{author}{Zegeer, C.V.}, \bibinfo{author}{Lyon, C.}, \bibinfo{year}{2010}.
\newblock \bibinfo{title}{Pedestrian safety prediction methodology for urban signalized intersections}.
\newblock \bibinfo{journal}{Transportation research record} \bibinfo{volume}{2198}, \bibinfo{pages}{65--74}.
\bibitem[{Truong and Somenahalli(2011)}]{truong2011using}
\bibinfo{author}{Truong, L.T.}, \bibinfo{author}{Somenahalli, S.V.}, \bibinfo{year}{2011}.
\newblock \bibinfo{title}{Using gis to identify pedestrian-vehicle crash hot spots and unsafe bus stops}.
\newblock \bibinfo{journal}{Journal of Public Transportation} \bibinfo{volume}{14}, \bibinfo{pages}{99--114}.
\bibitem[{Tubis et~al.(2021)Tubis, Skupie{\'n} and Rydlewski}]{tubis2021method}
\bibinfo{author}{Tubis, A.A.}, \bibinfo{author}{Skupie{\'n}, E.T.}, \bibinfo{author}{Rydlewski, M.}, \bibinfo{year}{2021}.
\newblock \bibinfo{title}{Method of assessing bus stops safety based on three groups of criteria}.
\newblock \bibinfo{journal}{Sustainability} \bibinfo{volume}{13}, \bibinfo{pages}{8275}.
\bibitem[{Tyndall(2021)}]{tyndall2021pedestrian}
\bibinfo{author}{Tyndall, J.}, \bibinfo{year}{2021}.
\newblock \bibinfo{title}{Pedestrian deaths and large vehicles}.
\newblock \bibinfo{journal}{Economics of Transportation} \bibinfo{volume}{26}, \bibinfo{pages}{100219}.
\bibitem[{Ulak et~al.(2021)Ulak, Kocatepe, Yazici, Ozguven and Kumar}]{ulak2021stop}
\bibinfo{author}{Ulak, M.B.}, \bibinfo{author}{Kocatepe, A.}, \bibinfo{author}{Yazici, A.}, \bibinfo{author}{Ozguven, E.E.}, \bibinfo{author}{Kumar, A.}, \bibinfo{year}{2021}.
\newblock \bibinfo{title}{A stop safety index to address pedestrian safety around bus stops}.
\newblock \bibinfo{journal}{Safety Science} \bibinfo{volume}{133}, \bibinfo{pages}{105017}.
\bibitem[{Vangala et~al.(2015)Vangala, Lord and Geedipally}]{vangala2015exploring}
\bibinfo{author}{Vangala, P.}, \bibinfo{author}{Lord, D.}, \bibinfo{author}{Geedipally, S.R.}, \bibinfo{year}{2015}.
\newblock \bibinfo{title}{Exploring the application of the negative binomial--generalized exponential model for analyzing traffic crash data with excess zeros}.
\newblock \bibinfo{journal}{Analytic methods in accident research} \bibinfo{volume}{7}, \bibinfo{pages}{29--36}.
\bibitem[{Walgren(1998)}]{walgren1998using}
\bibinfo{author}{Walgren, S.}, \bibinfo{year}{1998}.
\newblock \bibinfo{title}{Using geographic information system (gis) to analyze pedestrian accidents}, in: \bibinfo{booktitle}{68th Annual Meeting of the Institute of Transportation EngineersInstitute of Transportation Engineers (ITE)}.
\bibitem[{Washington et~al.(2020)Washington, Karlaftis, Mannering and Anastasopoulos}]{washington2020statistical}
\bibinfo{author}{Washington, S.}, \bibinfo{author}{Karlaftis, M.G.}, \bibinfo{author}{Mannering, F.}, \bibinfo{author}{Anastasopoulos, P.}, \bibinfo{year}{2020}.
\newblock \bibinfo{title}{Statistical and econometric methods for transportation data analysis}.
\newblock \bibinfo{publisher}{Chapman and Hall/CRC}.
\bibitem[{Williams(2012)}]{williams2012using}
\bibinfo{author}{Williams, R.}, \bibinfo{year}{2012}.
\newblock \bibinfo{title}{Using the margins command to estimate and interpret adjusted predictions and marginal effects}.
\newblock \bibinfo{journal}{The Stata Journal} \bibinfo{volume}{12}, \bibinfo{pages}{308--331}.
\bibitem[{Wretstrand et~al.(2014)Wretstrand, Holmberg and Berntman}]{wretstrand2014safety}
\bibinfo{author}{Wretstrand, A.}, \bibinfo{author}{Holmberg, B.}, \bibinfo{author}{Berntman, M.}, \bibinfo{year}{2014}.
\newblock \bibinfo{title}{Safety as a key performance indicator: Creating a safety culture for enhanced passenger safety, comfort, and accessibility}.
\newblock \bibinfo{journal}{Research in transportation economics} \bibinfo{volume}{48}, \bibinfo{pages}{109--115}.
\bibitem[{Xie et~al.(2017)Xie, Ozbay, Kurkcu and Yang}]{xie2017analysis}
\bibinfo{author}{Xie, K.}, \bibinfo{author}{Ozbay, K.}, \bibinfo{author}{Kurkcu, A.}, \bibinfo{author}{Yang, H.}, \bibinfo{year}{2017}.
\newblock \bibinfo{title}{Analysis of traffic crashes involving pedestrians using big data: Investigation of contributing factors and identification of hotspots}.
\newblock \bibinfo{journal}{Risk analysis} \bibinfo{volume}{37}, \bibinfo{pages}{1459--1476}.
\bibitem[{Yendra et~al.(2024)Yendra, Haworth and Watson-Brown}]{yendra2024comparison}
\bibinfo{author}{Yendra, D.}, \bibinfo{author}{Haworth, N.}, \bibinfo{author}{Watson-Brown, N.}, \bibinfo{year}{2024}.
\newblock \bibinfo{title}{A comparison of factors influencing the safety of pedestrians accessing bus stops in countries of differing income levels}.
\newblock \bibinfo{journal}{Accident Analysis \& Prevention} \bibinfo{volume}{207}, \bibinfo{pages}{107725}.
\bibitem[{Yu(2024)}]{yu2024impact}
\bibinfo{author}{Yu, C.Y.}, \bibinfo{year}{2024}.
\newblock \bibinfo{title}{The impact of built environments on pedestrian safety around bus stops}.
\newblock \bibinfo{journal}{Journal of Urban Affairs} \bibinfo{volume}{46}, \bibinfo{pages}{267--280}.
\bibitem[{Zahabi et~al.(2011)Zahabi, Strauss, Manaugh and Miranda-Moreno}]{zahabi2011estimating}
\bibinfo{author}{Zahabi, S.A.H.}, \bibinfo{author}{Strauss, J.}, \bibinfo{author}{Manaugh, K.}, \bibinfo{author}{Miranda-Moreno, L.F.}, \bibinfo{year}{2011}.
\newblock \bibinfo{title}{Estimating potential effect of speed limits, built environment, and other factors on severity of pedestrian and cyclist injuries in crashes}.
\newblock \bibinfo{journal}{Transportation research record} \bibinfo{volume}{2247}, \bibinfo{pages}{81--90}.
\bibitem[{Zamani and Ismail(2010)}]{zamani2010negative}
\bibinfo{author}{Zamani, H.}, \bibinfo{author}{Ismail, N.}, \bibinfo{year}{2010}.
\newblock \bibinfo{title}{Negative binomial-lindley distribution and its application}.
\newblock \bibinfo{journal}{Journal of mathematics and statistics} \bibinfo{volume}{6}, \bibinfo{pages}{4--9}.
\bibitem[{Zegeer and Bushell(2012)}]{zegeer2012pedestrian}
\bibinfo{author}{Zegeer, C.V.}, \bibinfo{author}{Bushell, M.}, \bibinfo{year}{2012}.
\newblock \bibinfo{title}{Pedestrian crash trends and potential countermeasures from around the world}.
\newblock \bibinfo{journal}{Accident Analysis \& Prevention} \bibinfo{volume}{44}, \bibinfo{pages}{3--11}.
\bibitem[{Zhang et~al.(2023)Zhang, Du, Shen and Ma}]{zhang2023revealing}
\bibinfo{author}{Zhang, C.}, \bibinfo{author}{Du, B.}, \bibinfo{author}{Shen, J.}, \bibinfo{author}{Ma, C.}, \bibinfo{year}{2023}.
\newblock \bibinfo{title}{Revealing safety impact of bus stops on passenger-cyclist interactions--evidence from nanjing, china}.
\newblock \bibinfo{journal}{Travel behaviour and society} \bibinfo{volume}{32}, \bibinfo{pages}{100578}.

\end{thebibliography}


\end{document}